\documentclass[aps,
prl,reprint,
superscriptaddress,
amsmath,amssymb,
longbibliography
]{revtex4-2}

\usepackage[normalem]{ulem}
\usepackage{multirow}
\usepackage{graphicx}
\usepackage{dcolumn}
\usepackage{bm}
\usepackage{orcidlink}
\usepackage{cancel}
\usepackage{xcolor}
\usepackage{float}
\usepackage{hyperref}
\hypersetup{colorlinks=true,
            citecolor = {cyan},
	        linkcolor = {blue},
            urlcolor  = {purple}
}
\usepackage[final]{pdfpages}
\usepackage{pgffor}
\usepackage{etoolbox}
\makeatletter
\AtBeginDocument{\let\LS@rot\@undefined}
\makeatother

\newcommand{\bz}{{\bf z}}

\newcommand{\kBT}{{k_{\rm B} T}}

\newcommand{\beq}{\begin{equation}}
\newcommand{\eeq}{\end{equation}}

\def\bal#1\eal{\begin{align}#1\end{align}}

\definecolor{rojo}{rgb}{1,0,0}
\definecolor{naranja}{rgb}{1,0.65098,0}
\definecolor{amarillo}{rgb}{1,1,0}
\definecolor{verde}{rgb}{0,1,0}
\definecolor{cian}{rgb}{0,1,1}
\definecolor{magent}{rgb}{1,0,1}
\definecolor{marron}{rgb}{0.898039,0.596078,0.4}
\definecolor{gris}{rgb}{0.698039,0.729412,0.733333}

\begin{document}

\title{Following the Committor Flow: A Data-Driven Discovery of Transition Pathways}

\author{Cheng Giuseppe Chen~\orcidlink{0000-0003-3553-4718}}
\affiliation{Laboratoire International Associ\'e Centre National de la Recherche Scientifique et University of Illinois at Urbana-Champaign, Unit\'e Mixte de Recherche n$^\circ$7019, Universit\'e de Lorraine, B.P. 70239, 54506 Vand\oe uvre-l\`es-Nancy cedex, France}

\author{Chenyu Tang~\orcidlink{0000-0002-6914-7348}}
\affiliation{Laboratoire International Associ\'e Centre National de la Recherche Scientifique et University of Illinois at Urbana-Champaign, Unit\'e Mixte de Recherche n$^\circ$7019, Universit\'e de Lorraine, B.P. 70239, 54506 Vand\oe uvre-l\`es-Nancy cedex, France}

\author{Alberto Meg\'ias~\orcidlink{0000-0002-7889-1312}}
\affiliation{Complex Systems Group and Department of Applied Mathematics, Universidad Polit\'ecnica de Madrid, Av. Juan de Herrera 6, E-28040 Madrid, Spain}

\author{Radu A. Talmazan~\orcidlink{0000-0001-6678-7801}}
\affiliation{Laboratoire International Associ\'e Centre National de la Recherche Scientifique et University of Illinois at Urbana-Champaign, Unit\'e Mixte de Recherche n$^\circ$7019, Universit\'e de Lorraine, B.P. 70239, 54506 Vand\oe uvre-l\`es-Nancy cedex, France}

\author{Sergio Contreras Arredondo~\orcidlink{0009-0006-0428-4962}}
\affiliation{Laboratoire International Associ\'e Centre National de la Recherche Scientifique et University of Illinois at Urbana-Champaign, Unit\'e Mixte de Recherche n$^\circ$7019, Universit\'e de Lorraine, B.P. 70239, 54506 Vand\oe uvre-l\`es-Nancy cedex, France}

\author{Beno\^\i t Roux~\orcidlink{0000-0002-5254-2712}}
\affiliation{Department of Biochemistry and Molecular Biology,  University of Chicago, Chicago, USA}
\affiliation{Department of Chemistry, University of Chicago, Chicago, USA}

\author{Christophe Chipot~\orcidlink{0000-0002-9122-1698}}
\email{chipot@illinois.edu}
\affiliation{Laboratoire International Associ\'e Centre National de la Recherche Scientifique et University of Illinois at Urbana-Champaign, Unit\'e Mixte de Recherche n$^\circ$7019, Universit\'e de Lorraine, B.P. 70239, 54506 Vand\oe uvre-l\`es-Nancy cedex, France}
\affiliation{Department of Biochemistry and Molecular Biology,  University of Chicago, Chicago, USA}
\affiliation{Theoretical and Computational Biophysics Group, Beckman Institute, and Department of Physics, 
University of Illinois at Urbana-Champaign, Urbana, USA}

\date{\today}
             
\begin{abstract}
    The discovery of transition pathways to unravel  distinct reaction mechanisms and, in general, rare events that occur in molecular systems is still a challenge. Recent advances have focused on analyzing the transition path ensemble using the committor probability, widely regarded as the most informative one-dimensional reaction coordinate. Consistency between transition pathways and the committor function is essential for accurate mechanistic insight. In this work, we propose an iterative framework to infer the committor and, subsequently, to identify the most relevant transition pathways. Starting from an initial guess for the transition path, we generate biased sampling from which we train a neural network to approximate the committor probability.  From this learned committor, we extract dominant transition channels as discretized strings lying on isocommittor surfaces. These pathways are then used to enhance sampling and iteratively refine both the committor and the transition paths until convergence. The resulting committor enables accurate estimation of the reaction rate constant. We demonstrate the effectiveness of our approach on benchmark systems, including a two-dimensional model potential, peptide conformational transitions, and a Diels–-Alder reaction.
\end{abstract}

\maketitle

\emph{Introduction---}Many complex  dynamical  systems exhibit metastability, a phenomenon wherein the system successively resides in some states for long periods of time. 
Understanding how such rare transitions take place is a fundamental problem across multiple disciplines, from chemistry and physics to biology and material science. In biomolecular systems, rare transitions are often associated with conformational changes, chemical reactions, or ligand binding processes, all essential to biological function~\cite{moradi2013,Lindorff-Larsen2011fold,ajaz_2011,yang2009}. More generally, similar metastable behavior is also seen in other complex abiological systems, such as magnetic-domain switching in ferromagnetic materials~\cite{Wulferding2017}, glassy systems~\cite{BK01}, or climate dynamics~\cite{lucente2022climate}. However, the properties of the underlying statistical-physical description of rare transitions between metastable states share common traits for such systems of very different nature. 

Central to understanding the mechanism underlying the function of biomolecular systems is a characterization of the different routes, or \emph{transition pathways}, taken by the system when interconverting between metastable states.
For instance, the insight gained from the knowledge of the relevant transition pathways can guide drug design, protein engineering, or the optimization of catalytic processes. A number of challenges, however, stand in the way of achieving this objective. In particular, the fact that the transitions at hand may be extremely rare makes their direct observation through simple brute-force simulations nearly impossible. From a practical standpoint, simulating these processes cannot be done successfully without some form of enhanced-sampling strategy~\cite{chen_enhancing_2022}. An additional challenge is the curse of dimensionality due to the large number of degrees of freedom at play. Usually, one expects that the necessary and sufficient information to understand and predict rare transitions is included within a subspace of lower dimension. Most practical computational strategies to tackle slow processes are designed to enhance the sampling of the associated degrees of freedom by leaning on the knowledge of a subspace of selected collective variables (CVs)~\cite{rogal_reaction_2021}. In this regard, information about the transition pathways between the metastable states within this subspace of CVs can be exploited to enhance sampling in a path-specific manner \cite{branduardi_b_2007}. On the basis of these observations, the overarching goal is to design an effective approach for discovering the relevant set of CVs alongside the most likely set of transition pathways  that underlie the slow processes of interest. 

A given transition pathway connecting two long-lived metastable states $A$ and $B$ can be parametrically represented as a \emph{reaction tube} of a given width centered on a curvilinear line embedded in the subspace of the CVs \cite{Vanden-Eijnden-2010}. These formal constructs appear also in the context of the finite-temperature string method in CV space 
\cite{e_finite_2005}, 
the string method in collective variables yielding the minimum free energy pathway (MFEP)
\cite{maragliano2006string}, and the string method with swarms-of-trajectories supplying a zero-drift pathway (ZDP) \cite{pan_finding_2008}. For completeness, it should be mentioned that the MFEP can be inferred from a complete mapping of the free-energy landscape \cite{fu2020finding}. In the last decade, the development of transition path theory (TPT) has established a theoretical framework to greatly expand the concept of transition pathways. One of the central ingredients of TPT is the concept of forward committor probability, $q$, defined as the probability that a trajectory initiated at some configuration, $\bz_0$, will ultimately reach state $B$ before state $A$~\cite{VT05}, i.e.,
\begin{equation}\label{eq:def_q}
    q(\bz_0) = P\big(\tau_B(\bz_0)<\tau_A(\bz_0)\big),
\end{equation}
where $\displaystyle \tau_S = \inf_t \{ \bz(t|\bz_0)\in S\}$ is the hitting time of $S\subset \Omega$ with initial condition $\bz_0$, with $\bz(t|\bz_0)$ being the trajectory point at time $t$ from an initial condition $\bz_0$.  Importantly, knowledge of the committor, $q$, enables a formal expression of all reactive events from $A$ to $B$, providing key information for the definition and discovery of dynamically meaningful transition pathways. In this sense, the function $q$ can serve as the best surrogate for the reaction coordinate (RC) in the subspace of CVs \cite{K18}. The committor can also be expressed in terms of a functional minimization of the steady-state reactive flux, $J_{AB}[q;\tau]$, which can be written in terms of a time-correlation function~\cite{roux2022transition}, i.e.,
\begin{eqnarray}\label{eq:Jab}
    J_{AB}[q;\tau] & = & \frac{C[q;\tau]}{\tau},
    \nonumber
    \\
    C[q;\tau] & = & \frac{1}{2}\left\langle(q(\tau)-q(0))^2 \right\rangle
\end{eqnarray}
where $\tau$ is the time lag that guarantees a Markovian description of the dynamics, and $q(t)\equiv q(\bz(t))$ is the value taken by the committor at time $t$ of the trajectory. Said differently, $q$, as defined in Eq.~\eqref{eq:def_q}, is the function that minimizes $J_{AB}[q;\tau]$ for a given $\tau$.  These considerations together with the variational principle led to the concept of committor-consistent string (CCS) \cite{he_committor-consistent_2022}. The latter is generally not coincident with the MFEP or the ZDP. The theoretical framework from TPT, when combined with the power of artificial neural networks (ANN), allows us to develop tools designed to learn the committor probability from dynamical information. An example of these tools is provided by the variational committor networks (VCNs)~\cite{chen_discovering_2023}, 
based on Eq. \eqref{eq:Jab}. 

One of the main features of the committor function is that the different transition mechanisms can be found in the vicinity of the so-called separatrix, defined as the isocommittor hyperplane at $q=1/2$. Hence, this hyperplane contains all configurations that have equal probability of evolving towards $A$ or $B$. In the Markovian diffusive limit, the tangent to the curvilinear CCS transition pathway should follow the local gradient of the function $q$ in the neighborhood of the separatrix \cite{berezhkovskii_one-dimensional_2005, roux2022transition}. This idea also aligns with recent insights from information thermodynamics of transition-state ensembles the optimal RC should follow the gradient of the committor \cite{LS22}. These fundamental observations form the cornerstone of this work. 

Both the determination of the optimal pathways and learning of the committor depend on prior knowledge of the underlying dynamics. This requirement can be addressed via iterative approaches, whereby initial sampling is necessary to prime the optimization workflow. Improvement of the sampling along the model RC leverages the knowledge of the inferred committor. An illustration of this feedback loop is furnished in Ref.~\cite{MCCCRC25}, where the committor is learned concomitantly with its corresponding CCS. However, in complex molecular systems, there is no guarantee that the transition at hand can be modeled by a single CCS. In reality, multiple relevant pathways often form the transition-path ensemble. This plurality of routes makes the full characterization of the underlying reaction mechanisms appreciably more intricate. In this Letter, we introduce a methodology geared towards the discovery of multiple committor-guided transition pathways, or reaction tubes. This methodology consists of an iterative approach to predict and refine the committor based on multiple simulations performed along the pathways determined self-consistently. In addition, we show that the transition kinetics can be accurately modeled from the converged ensemble of pathways.

\begin{figure*}
    \centering
    \includegraphics[width=\linewidth]{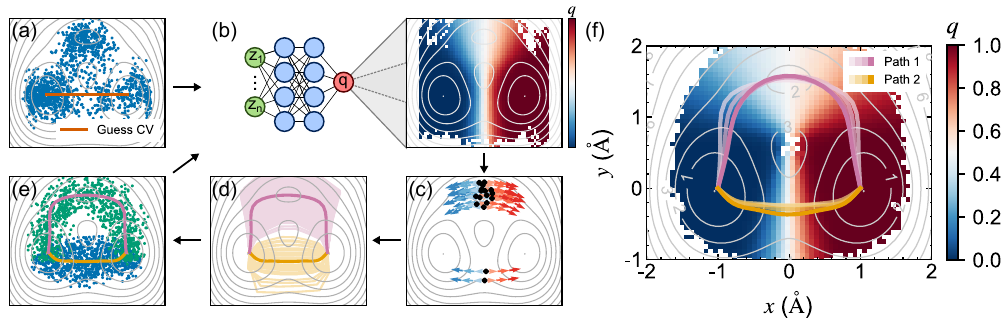}

    \caption{(a-e) Schematic representation of the steps of our iterative approach: (a) Initial biased sampling, (b) learn the committor, (c) forward and backward shooting of strings from the separatrix, (d) clustering of the gradient-guided strings and calculating the average strings, (e) biased sampling along the new pathways. (f) Converged CCSs and learned committor values $q$ corresponding to the two transition pathways of the triple-well potential. Strings from the iterations are also included, with a color gradient indicating the progression of the iterative procedure.}
    \label{fig:fig1}
\end{figure*}

\emph{Methodology---}Let us assume that the pathway can be parameterized in terms of a discretized string of $m$ images in the CV subspace $\Omega \subseteq \mathbb{T}^{n_1}\times\mathbb{R}^{n_2}$, $\{\widetilde{\bz}_i\}_{i=0}^{m-1}$, where $\mathbb{T}^{n_1}\times\mathbb{R}^{n_2}$ is the total CV subspace for, in general, $n_1$ periodic and $n_2$ nonperiodic CVs. This discretized string will be henceforth referred to as \emph{string pathway}. Under these premises, we introduce an iterative loop designed to reveal multiple transition pathways, drawing inspiration from feedback-based methodologies commonly employed to estimate the committor probability. The insight gained from each iteration is leveraged to enhance sampling, thereby refining progressively the results until convergence is achieved. This process is outlined in the following workflow:
\begin{itemize}
    \item[(a)] Generate biased sampling based on an initial guess of the CVs.
    \item[(b)] Use the VCN to learn the committor function over a predefined set of CVs, $\bz$, in the $\Omega$ subspace, not necessarily identical to that of the previous step.
    \item[(c)] Identify all states for which the committor value lies within the interval $q \in [1/2 - \varepsilon, 1/2 + \varepsilon]$, with $2\varepsilon \ll 1$. From these states, propagate in $\Omega$ deterministic string pathways along the gradient of the committor, $\nabla_{\bz} q$, backward and forward, so as to reach state $A$ and state $B$, respectively. Ultimately, the generated half pathways are mended together to form a continuous one. 
    \item[(d)] Cluster the resulting pathways and compute a weighted average for each cluster to obtain representative string pathways.
    \item[(e)] Perform biased simulations along CVs defined by each representative string pathway, e.g., the discretized path-collective variable (PCV)~\cite{branduardi_b_2007}. If convergence is achieved, terminate the workflow. Otherwise, return to step (b).
\end{itemize}

\begin{figure*}
    \centering
    \includegraphics[width=\linewidth]{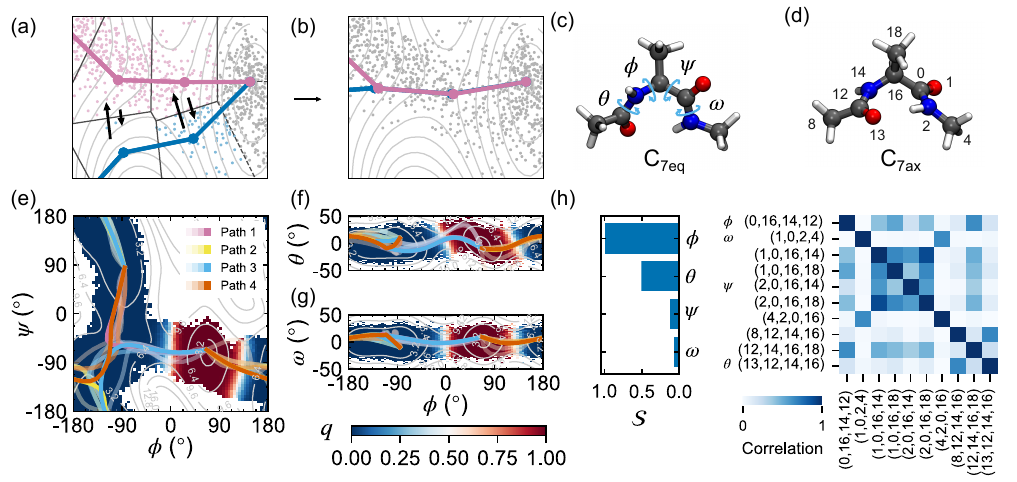}
    \caption{(a-b) Schematic representation of the path merging to identify overlapping paths: (a) Exchange in the sampled trajectory between the Voronoi tiles defined by the string images, (b) new string images from weighted averages. (c-d) Rendered representation of NANMA in (c) the C$_{\mathrm{7eq}}$ and (d) C$_{\mathrm{7ax}}$ conformations. Carbon atoms are depicted in grey, hydrogen in white, oxygen in red, and nitrogen in blue. The dihedral angles, $\phi$, $\psi$, $\theta$, and $\omega$, are highlighted in (c), while the labels of the heavy atoms are shown in (d). (e-g) Converged CCSs and learned committor, $q$, corresponding to the relevant isomerization pathways of NANMA, projected onto the (e) ($\phi$, $\psi$), (f) ($\phi$, $\theta$), and (g) ($\phi$, $\omega$) subspaces. The reference free-energy landscape is shown as gray contour lines, with the values expressed in kcal/mol. (h) Right: Pearson correlation matrix between all the dihedral angles of NANMA. Left: Nondimensional sensitivity, $\mathcal{S}$, of the model with respect to the CVs belonging to $\Omega$, selected from the correlation analysis.}
    \label{fig:nanma}
\end{figure*}

The various steps of the methodology are gathered in Fig.~\ref{fig:fig1}(a--e). It is noteworthy that in order to guarantee the smoothness of $\nabla_{\bz} q$, the original VCN introduced in Ref.~\cite{chen_discovering_2023} has been slightly tailored, as described in the Supplemental Material (SM). As stated previously, the objective of the iterative procedure is to progressively refine the learned committor function, thereby improving the identification of the transition pathways. In the spirit of Refs.~\cite{he_committor-consistent_2022} and \cite{MCCCRC25}, the string pathways are defined to be perpendicular to the separatrix hyperplane, and, consequently, aligned with the gradient of the committor. By extension, we assume that these pathways strictly follow $\nabla_{\bz} q$ (forward and backward) at each point between the separatrix and the basins. This construction addresses the degeneracy problem often encountered when characterizing transition pathways in complex molecular processes~\cite{YS25,MSPR16}. Specifically, to prevent double-counting of transitions occurring at different spatial locations, we implement a post-processing step during the determination of the representative string pathways. Specifically, in step~(d) of the workflow, we analyze the exchange between different averaged strings by performing a Voronoi tessellation of $\Omega$, wherein the string images are the centroids of each cell. Cells between which significant exchange is measured are merged into a single tile, thus avoiding artificial splitting and revealing the actual reaction tubes for the molecular process at hand. In practice, this procedure described schematically in Fig.~\ref{fig:nanma}(a--b), eliminates the risk of redundant sampling of effectively identical reaction tubes.


The curvilinear strings of images used in the enhanced sampling PCV calculations serve as proxies to explore the reaction tubes delineating the transition pathways associated with the different reaction mechanisms~\cite{branduardi_b_2007}. 

Consequently, small changes in the string pathways between successive iterations are expected to not significantly affect sampling. It follows that the convergence criteria should rely on the quality and consistency of sampling rather than on the precise shape of the string pathways. We quantify the consistency of sampling using the Kullback--Leibler divergence (KLD), $\mathcal{D}$, based on the spatial distribution of samples throughout $\Omega$. After computing the representative discretized string pathways—--applying the merging criteria based on exchanges between Voronoi cells—--we obtain a set of $n_s$ images, $\{\widetilde{\bz}_i\}_{i=1}^{n_s}$. These images define a partition of $\Omega$ into Voronoi cells, $\{C_i\}_{i=1}^{n_s}$, where each cell is given by $C_i = \left\{ \bz \in \Omega \mid d(\bz, \widetilde{\bz}_i) \leq d(\bz, \widetilde{\bz}_j) \quad \forall j \neq i \right\}$, with $d$ denoting a distance. In practice, this distance coincides with the Euclidean one, taking into account periodicity wherever needed. From this partition, we infer the sampling probability for each cell $C_i$ at iteration $k$, referred to as $p^{(k)}(C_i)$, and compare it to the probability from the previous iteration, $p^{(k-1)}(C_i)$. Next, the KLD at the $k$-th iteration can be defined as,
\begin{equation}
    \mathcal{D}^{(k)} = \sum_{i=1}^{n_s} p^{(k)}(C_i)\ln\left( \frac{p^{(k)}(C_i)}{p^{(k-1)}(C_i)}\right).
\end{equation}
If the iterative process has converged, the simulations performed under identical conditions should yield consistent spatial sampling probabilities. Consequently, one expects ${\cal D}^{(k)}$ to decrease progressively at each iteration $k$. Similar KLD measures have been employed previously to detect differences in spatial distributions in a variety of contexts~\cite{BRGD19,MS21}.
Once convergence of the sampling has been established, our  ultimate aim is to compute the rate constant for the transition at hand by recovering the unbiased flux, $J_{AB}$. To achieve this goal, we generate a series of unbiased trajectories from random configurations corresponding to $\{{\bz}_{i}\}$ taken from $n$ separate PCV trajectories. 
The time-correlation function of Eq.~\eqref{eq:Jab} can be calculated as a weighted average over initial conditions, ${\bz}_i$, distributed uniformly in the CV subspace, $\Omega$, of these unbiased trajectories~\cite{harris2025membrane},
\begin{equation}
    C[q;t] =    \frac{\displaystyle \sum_i R({\bz}_i) \ c_{\bz_i}[q;t]}{\displaystyle \sum_j R({\bz}_j)},
\label{eq:C[q;t]}
\end{equation}
where $R({\bz}_i)$ is a reweighting factor for point $\bz_i$ randomly picked from the totality of the biased PCV data, and used as the initial condition to generate a short unbiased trajectory of $m_t$ steps, $\{\bz(j\Delta t | \bz_i) \}_{j=0}^{m_t-1}$, with time step $\Delta t$, which is then used to calculate the conditional time-correlation function of the learned committor at time $t$, 
\begin{equation} 
    c_{\bz_i}[q;t] = \displaystyle \frac{1}{m_t}\sum_{j=0}^{m_t-1} \Big(q\big(\bz (t+j\Delta t | \bz_i)\big)- q\big(\bz (j\Delta t | \bz_i) \big)\Big)^2.
\end{equation}
The reweighting factor for point $\bz_i$ is defined as,
\begin{equation}\label{eq:R}
     R({\bz}_i)  =  \left(\displaystyle \sum_j n_j \ e^{-\beta\left(w_j(\bz_i)-F_j\right)}\right)^{-1},
\end{equation}
where $\beta=1/\kBT$. Eq.~\eqref{eq:R} is determined from a bin-less version of the weighted histogram analysis method (WHAM) \cite{Kumar1992}, leveraging the data accrued in the $n$ separate PCV calculations,
\begin{equation}
\left\{
\begin{array}{rll}
         F_j  & = & \displaystyle -\frac{1}{\beta} \ln \int {\rm d}{\bz} \ e^{-\beta w_j({\bz})}\langle \rho({\bz})\rangle, \\ [0,8cm]
    \langle \rho(\bz)\rangle & = & \frac{\displaystyle \sum_j n_j \langle \rho(\bz)\rangle_{\rm b}^{(j)}}{\displaystyle\sum_k n_k e^{-\beta \left(w_k(\bz)-F_k\right)}},
\end{array} 
\right.
\end{equation}
where $n_j$ and $w_j({\bf z})$ are the total number of data points and the converged biasing potential (see SM) from the $j$-th PCV calculation, respectively, and $F_j$ is the free-energy offset determined in a self-consistent fashion by the $j$-th PCV calculation via the WHAM equations from the biased distribution, $\langle \rho(\bz)\rangle_{\rm b}^{(j)}$ (generated by the $j$-th PCV calculation). Here, $\langle \rho(\bz)\rangle$ formally stands for the complete unbiased equilibrium distribution in the CV subspace, $\Omega$, although it is not explicitly generated in the current bin-less treatment.
For more details about the derivation of this methodology, the reader is referred to the SM. Finally, recalling Eq.~\eqref{eq:Jab}, the steady-state reactive flux can be recovered as $\displaystyle J_{AB} = \lim_{t\to \tau_q} {\rm d}C[q;t]/{\rm d}t$, where $\tau_q$ corresponds to a relaxation time, allowing the rate constant to be estimated from $k_{AB} = J_{AB} / p_A$, where $p_A=P(\bz\in A)$ is the probability to be in the reactant basin $A$ approximated from event counting. 

\emph{Results---}As a first illustration, we assessed our methodology with a toy model---the two-dimensional triple-well potential~\cite{metzner_2006,metzner_2008} (see Fig.~\ref{fig:fig1}f). Through the aforementioned iterative process, starting from an initial PCV sampling at 300~K along a guess straight path connecting the two main basins, we were able to identify the two expected string pathways: An upper one crossing a local minimum, and a lower one, favored at higher temperature, requiring a higher energy barrier to be overcome~\cite{MCCCRC25}.

Next, we investigated a simple molecular system, $N$--acetyl--$N$$^\prime$--methylalanylamide (NANMA)---often called dialanine or alanine dipeptide---in vacuum. The isomerization of NANMA between the C$_{\mathrm{7eq}}$ and C$_{\mathrm{7ax}}$ conformations (see Fig.~\ref{fig:nanma}c--d) has been extensively used as a benchmark to validate new methodologies aimed at sampling rare events~\cite{RAT05,HFCK10,bolhuis_reaction_2000}. In this work, we chose the backbone dihedral angles $\phi$, $\psi$, $\theta$ and $\omega$ as the CVs (see Fig.~\ref{fig:nanma}c), in line with previous studies of this system~\cite{chen_discovering_2023, kang2024Parrinello, Trizio2025, MCCCRC25}. 
Most notably, we were able to obtain the same set of CVs using a statistical analysis. To do so, we started enumerating all possible dihedral angles of NANMA as CV candidates. Then, we calculated the Pearson correlation matrix and pruned the redundant CVs. Among each group of correlated dihedral angles, only the one with the highest importance---defined as the sum of the correlation coefficients involving the CV at hand---was kept (see Fig.~\ref{fig:nanma}h). Moreover, the nondimensional sensitivity, $\mathcal{S}_k$, of the model with respect to the uncorrelated $k$-th CV, defined as, $\displaystyle{\mathcal{S}(z_i) = s_i/\max_k s_k }$, with $s_i = \left\langle\left| \partial_{z_i} q(\bz) \right|\right\rangle_{\bz \in \Omega} \equiv \left\langle\left| \partial q(\bz) / \partial z_i \right|\right\rangle_{\bz \in \Omega}$, can be computed to further discriminate the relevant CVs by ranking their importance in the dynamics of the process. These results, while preliminary, are suggestive of a general framework for the definition of $\Omega$.
To prime the iterative process, we ran a simulation biasing the two RMSDs with respect to C$_{\mathrm{7eq}}$ and C$_{\mathrm{7ax}}$. Our proposed approach was able to correctly identify all the relevant transition pathways~\cite{bolhuis_reaction_2000,branduardi_b_2007,Lee2017} (see Fig.~\ref{fig:nanma}e--g), while other similar data-driven committor-based schemes fail to achieve this objective, managing to single out only one of the possible paths (corresponding to Path 1 in Fig~\ref{fig:nanma}e--g)~\cite{kang2024Parrinello, Trizio2025, MCCCRC25}.

\begin{figure}
    \centering
    \includegraphics[width=\linewidth]{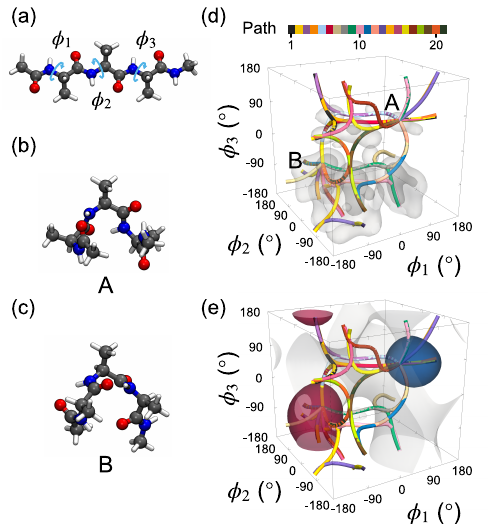}
    \caption{(a-c) Graphical representation of trialanine: (a) definition of the $\phi_1$, $\phi_2$, and $\phi_3$ dihedral angles, (b) conformation A, and (c) conformation B. C atoms are depicted in grey, H in white, O in red, and N in blue. (d-c) Converged CCSs with (d) the reference free-energy landscape (grey isosurfaces corresponding to a free energy of 3.3 and 6.5 kcal/mol) and (e) the learned committor $q$ and its isosurfaces corresponding to $q = 0$ (blue), $q = 0.5$ (grey), and $q = 1$ (red).}
    \label{fig:triala}
\end{figure}

To further appraise the ability of our methodology to handle multiple, alternate and interconnected pathways, we tackled the challenging case of the isomerization of trialanine in vacuum, using ($\phi_1$, $\phi_2$, $\phi_3$) as the CVs~\cite{Tiwary2017Predicting, chen_companion_2022} (see Fig.~\ref{fig:triala}a). To prime the iterative process, we ran a simulation biasing two RMSDs with respect to state $A$ defined by (60$^\circ$, $-$70$^\circ$, 60$^\circ$) and state $B$ defined by ($-$70$^\circ$, 60$^\circ$, $-$70$^\circ$) (see Fig.~\ref{fig:triala}b--c). The rugged free-energy landscape underlying isomerization~\cite{fu_taming_2019} suggests that transiting from $A$ to $B$ can occur through a variety of pathways, passing through multiple intermediates. A previous study showcased a computational strategy based on the string method with swarms of trajectories~\cite{pan_finding_2008}, from which the authors were able to obtain five relevant transition pathways~\cite{chen_companion_2022}. Our findings not only confirm the results of Chen et al., barring the slowest, less relevant path, but also bring to light a total of 21 different string pathways connecting $A$ and $B$ (see Fig.~\ref{fig:triala}d,e, and Table S2 in the SM), thus rendering a complete picture of the conformational equilibrium dynamics.

\begin{figure}[ht!]
    \centering
    \includegraphics[width=\linewidth]{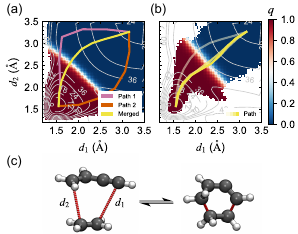}
    \caption{Committor and strings of the Diels--Alder reaction before and after the Voronoi tesselation merging (a). The converged CCS and committor after three iterations (b). The reactants and product are shown in panel c.}
    \label{fig:DA_paths}
\end{figure}

As a final illustration, we considered the Diels--Alder reaction of ethylene with vinyl--acetylene, depicted in Fig.~\ref{fig:DA_paths}c. This reaction showcases the importance of the path-merging algorithm described above. The initial sampling was obtained by biasing along the $d_1$ distance shown in Fig.~\ref{fig:DA_paths}c. In the first iteration of the workflow, two distinct pathways were determined (see Fig.~\ref{fig:DA_paths}a), which were subsequently merged into a single string pathway (see Fig.~\ref{fig:DA_paths}b), leading to the formation of the product. This result aligns with the two-dimensional quantum-chemical free-energy landscape mapped for this reaction, suggestive that the two initial pathways were, in fact, alternative routes that pertain to the same very broad reaction tube.

The convergence of all calculations reported herein is assessed based on the progressive decrease of ${\cal D}$ as a function of the iteration, as detailed in the SM. At the end of each iterative process, short unbiased simulations were performed from a subset of configurations extracted from the PCV calculation of the last iteration. These short unbiased trajectories were then used to learn the final committor function, from which the rate constants, $k_{AB}$, associated with the molecular processes of interest were estimated. The estimates reported in Table~\ref{tab:table1} correspond to a 99\% confidence interval, and are consistent with values reported using alternate methodologies. The detail of these computations can be found in the SM.

\begin{table}
\caption{\label{tab:table1}%
Transition rates $k_{AB}$ for the triple-well potential, NANMA and trialanine isomerization, and the Diels--Alder reaction.}
\begin{ruledtabular}
\begin{tabular}{lrl}
\multirow{2}{*}{System} & \multicolumn{2}{c}{Transition rate, $k_{AB}$ (ps$^{-1}$)}\\[0.1cm]
\cline{2-3}
\\[-0.2cm]
 & This work \hspace*{0.6cm} & Reference \\
\hline
\\[-0.2cm]
Triple-well & $(1.450 \pm 0.010) \times 10^{-2}$ & $6.44 \times 10^{-2}$~\cite{MCCCRC25}\\
NANMA  & $(3.0 \pm 0.7)\times10^{-6}$ & $1.34 \times 10^{-5}$~\cite{chen_discovering_2023}\\
Trialanine & $(1.8 \pm 0.4) \times 10^{-5}$ & $3.62 \times 10^{-5}$~\cite{chen_companion_2022}\\
Diels--Alder reaction & $(2.3\pm 1.0)\times10^{-4}$ & $1.03\times10^{-3}$~\cite{ajaz_2011} \\
\end{tabular}
\end{ruledtabular}
\end{table}

\emph{Conclusion---}In this Letter, we have introduced a novel approach for the iterative discovery of committor-guided string pathways, used in turn to enhance the sampling of rare events. The key idea lies in the construction of string pathways that follow the gradient of the committor function \cite{berezhkovskii_one-dimensional_2005,LS22}, initiated from configurations sampled at the separatrix, that is at $q$ = 0.5. This committor-consistent framework allows the underlying dynamics of the molecular process at hand to be faithfully rendered. By clustering the ensemble of curvilinear lines determined in the $\Omega$ subspace of CVs, we identify the dominant transition pathways that characterize the process of interest. In systems evolving on rugged free-energy landscapes---as often observed in high-dimensional biomolecular dynamics---multiple competing pathways may coexist. Our approach resolves the inherent degeneracy associated with committor-based descriptions by leveraging the committor directly as a CV, thereby enabling efficient sampling across all relevant pathways. Moreover, by merging string pathways based on transitions between Voronoi cells in the $\Omega$ subspace, we recover the correct dynamical behavior and uncover overlapping reaction tubes. Our methodology is general---it is agnostic to the specific means used to estimate the committor or to perform sampling, and is thus compatible with a broad range of simulation strategies aimed at exploring rare events. One remaining limitation lies in the definition of the $\Omega$ subspace. As a possible route to address this limitation, we propose a preliminary framework for selecting meaningful CVs from a large pool of potential candidates based on the gradient of the committor. Finally, the identification of overlapping reaction tubes opens the door to further efficiency gains. In particular, decomposing string pathways into unique segments and employing multiple-copy and replica-exchange strategies may help reduce redundant sampling in future implementations.

\emph{Acknowledgements---}C.C. acknowledges the European Research Council (project 101097272 ``MilliInMicro''), the Université de Lorraine through its Lorraine Université d'Excellence initiative, the Région Grand-Est (project ``Respire''), and the Agence Nationale de la Recherche under France 2030 (contract ANR-22-PEBB-0009) for support in the context of the MAMABIO project (B-BEST PEPR). A.M. is sincerely thankful for the support provided by Universidad Polit\'ecnica de Madrid through the ``Programa Propio de Investigaci\'on'' grant EST-PDI-25-C0ON11-36-6WBGZ8, as well as the hospitality of the Université de Lorraine during his stay.

\clearpage


\section*{Supplementary material}

# Supplemental Material:
# Following the Committor Flow: A Data-Driven Discovery of Transition Pathways

Cheng Giuseppe Chen ●,[1] Chenyu Tang ●,[1] Alberto Megías ●,[2] Radu A. Talmazan ●,[1]
Sergio Contreras Arredondo ●,[1] Benoît Roux ●,[3,4] and Christophe Chipot ●[1,3,5,*]

[1] *Laboratoire International Associé Centre National de la Recherche Scientifique et University of Illinois at Urbana-Champaign,*
*Unité Mixte de Recherche n° 7019, Université de Lorraine,*
*B.P. 70239, 54506 Vandœuvre-lès-Nancy cedex, France*
[2] *Complex Systems Group and Department of Applied Mathematics,*
*Universidad Politécnica de Madrid, Av. Juan de Herrera 6, E-28040 Madrid, Spain*
[3] *Department of Biochemistry and Molecular Biology, University of Chicago, Chicago, USA*
[4] *Department of Chemistry, University of Chicago, Chicago, USA*
[5] *Theoretical and Computational Biophysics Group, Beckman Institute,*
*and Department of Physics, University of Illinois at Urbana-Champaign, Urbana, USA*

## CONTENTS

## S1. SIMULATION AND LEARNING DETAILS

All molecular dynamics (MD) simulations were performed using the MD package NAMD 3.0.1 [1] and the Colvars library [2]. All the biased simulations were run using well-tempererd metadynamics extended adaptive biasing force (WTM-eABF) algorithm [3], with the exception of the study of the Diels–Alder reaction, where multiple-walker (MW) WTM-eABF [3, 4] was employed instead. All the MD simulations where performed at a constant temperature of 300 K through the use of the Langevin thermostat, with an integration timestep of 0.5 fs. The covalent bonds involving hydrogen were constrained using the RATTLE algorithm [5].

### A. The variational committor network

The variational committor network (VCN), as presented in Ref. [6], consists of 4 layers of 32 nodes each with ELU [7] as activation function for each. This artificial neural network (ANN) is founded on the minimization of the

———
* chipot@illinois.edu

following loss function,

$$\mathcal{L}_{\text{VCN}}[q_\omega] = 2C[q_\omega; \tau],\tag{S1}$$

which is based on the variational principle introduced in Refs. [8, 9], with $q_\omega$ computed as

$$q_\omega(\mathbf{z}) = \begin{cases} 0, & \text{if } \mathbf{z} \in A, \\ F_\omega(\mathbf{z}), & \text{if } \mathbf{z} \in (A \cup B)^c, \\ 1, & \text{if } \mathbf{z} \in B, \end{cases}\tag{S2}$$

where $F_\omega$ is the scalar function learned by the VCN, and $\omega$ refers to the parameters of the ANN.

However, Eq. (S2) does not guarantee the boundedness of the output within the interval $[0, 1]$, as illustrated in the original paper (Ref. [6]), where regions with values outside this interval are shown in grey. To address this issue, we added a final layer consisting of a single node with a sigmoid activation function, $\sigma(x) := 1/(1 + e^{-x})$, which strictly constrains the output to the $[0, 1]$ range.

Additionally, following the approach in Ref. [10], we included a loss term on the restraints to enforce a continuous mapping at the boundaries between $(A \cup B)^c$ and the basins. The total loss function is therefore given by

$$\mathcal{L}[q_\omega] = 2C[q_\omega; \tau] + \lambda \left( |F_\omega(\mathbf{z})|^2_{\mathbf{z} \in A} + |F_\omega(\mathbf{z}) - 1|^2_{\mathbf{z} \in B} \right),\tag{S3}$$

with $\lambda \in \mathbb{R}^+$ a constant set a priori to control the boundary loss.

Finally, we kept the same number of layers of the original VCN and extended their number of nodes to 64 nodes to improve the learning process.

### B. Triple-well potential

The triple-well potential studied in this work is defined as,

$$V(x, y) = 3e^{-x^2 - \left(y - \frac{1}{3}\right)^2} - 3e^{-x^2 - \left(y - \frac{5}{3}\right)^2} - 5e^{-(x-1)^2 - y^2} - 5e^{-(x+1)^2 - y^2} + \frac{2}{10}x^4 + \frac{2}{10}\left(y - \frac{1}{3}\right)^4.\tag{S4}$$

For the MD simulations, a damping coefficient of 100 ps$^{-1}$ was used for the Langevin thermostat. For the WTM-eABF algorithm, the PCV was sampled between 0.1 and 0.9, with a bin width of 0.01. Harmonic walls with a force constant of 1 kcal/mol were placed to restrain the system within a PCV value of 0.1 and 0.9. After the initial sampling of the iterative method, an additional harmonic wall of force constant 30 kcal/(mol·Å$^4$) was placed to restrain the value of the orthogonal PCV, $\zeta$ (see its definition in Eq. (S6) [11]), below 0.2 Å$^2$. The biasing force was applied after 100 samples were collected in each bin, using an extended fluctuation of 0.01, an extended time constant of 200 fs, a bias temperature of 1,000 K depositing the Gaussian biasing potentials with an initial height of 0.1 kcal/mol, a width of 0.03 Å and a frequency of 1 every 1,000 steps. For the initial sampling, we ran a 10 ns PCV simulation along the straight line connecting the two basins [$A = (-1, 0)$ and $B = (1, 0)$]. For the following three iterations, we ran a 5 ns PCV simulation for each CCS. Only the last 8 ns from the initial sampling and the last 4 ns for the other iterations were used to train the VCN. After the final iteration, 2,000 unbiased simulations of 30 ps each were performed starting from conformations sampled in the PCV simulations along the converged CCSs.

For the training of the VCN, the time-correlation function of the committor $q$ was calculated using a time-lag $\tau$ of 100 fs. The trainings were performed with a maximum number of epochs of 5,000, with an early stopping set at 20 epochs.

The gradient-guided paths were generated at each iteration by randomly selecting 500 sampled points close to the separatrix as the initial points. Then, they were clustered using hierarchical clustering, based on the RMSD between different strings, with a cut-off height of 0.6 Å.

### C. NANMA isomerization in vacuum

NANMA in vacuum was simulated using the CHARMM22 force field [12]. A damping coefficient of 10 ps$^{-1}$ was used for the Langevin thermostat. For the initial sampling, we performed a 5 ns WTM-eABF simulation using, as CVs, the two root mean square displacements (RMSDs) of the heavy atoms with respect to conformations C$_{7\text{eq}}$ and C$_{7\text{ax}}$. The CVs were sampled in the range between 0.1 and 1.8 Å with a bin width of 0.05 Å. Harmonic walls of force constant 1 kcal/(mol·Å$^2$) were set to restrain the RMSDs within 0.1 and 1.8 Å. The regions in the CV subspace

that remained unsampled after 1 ns were masked for the rest of the simulation in order to avoid nonphysical regions (e.g., where both RMSDs are sufficiently small). The biasing force was applied after 5,000 samples were collected in each bin, using an extended fluctuation of 0.01 Å, an extended time constant of 200 fs, a bias temperature of 2,000 K depositing the Gaussian biasing potentials with an initial height of 0.1 kcal/mol, a width of 0.1 Å and a frequency of 1 every 1,000 steps. For the following iterations, we performed 5 ns PCV simulations for each CCS. The PCV was sampled between 0.05 and 0.95, with a bin width of 0.01. Harmonic walls were set to restrain the PCV within 0.05 and 0.95 using force constants of 1 kcal/mol, and $\zeta$ below 600 degree[2] using a force constant of 0.001 kcal/mol degree[4]. The same WTM-eABF parameters were used for the eABF portion of the algorithm. For the WTM part, we employed a bias temperature of 1,000 K, an initial hill height of 2 kcal/mol, a width of 0.02 and a frequency of 1 every 1,000 steps. After the final iteration, 4,000 unbiased simulations of 10 ps each were performed starting from conformations sampled in the PCV simulations along the converged CCSs.

For the training of the VCN, only the last 4 ns of each biased simulation was used. The time-correlation function of the committor $q$ was calculated using a time-lag $\tau$ of 100 fs. The VCN consisted of 4 layers with 64 nodes each, and the trainings were performed with a maximum number of epochs of 5,000, with an early stopping set at 20 epochs.

For the initial sampling, 300 initial points were used to generate the gradient-guided pathways. For the following iterations, 200 initial points were used except for the final iteration, where 1,000 points were selected instead. The string pathways were then clustered using hierarchical clustering using a cut-off height of 0.3°.

### D. Trialanine conformational equilibria in vacuum

Trialanine in vacuum was simulated using the AMBER ff14SB force field [13] and a damping coefficient of 1 ps$^{-1}$. For the initial sampling, we performed a 120 ns WTM-eABF simulation using, as CVs, the two RMSDs of the heavy-atoms positions with respect to those of conformations $A$ ($\phi_1 = 60°$, $\phi_2 = -70°$, $\phi_3 = 60°$) and $B$ ($\phi_1 = -70°$, $\phi_2 = 60°$, $\phi_3 = -70°$). The CVs were sampled in the range between 0.05 and 5.05 Å with a bin width of 0.1 Å. Harmonic walls of force constant 100 kcal/(mol·Å²) were set to restrain the RMSDs below 5.5 Å. The regions in the CV-subspace that remained unsampled after 1 ns were masked for the rest of the simulation in order to avoid nonphysical regions. The biasing force was applied after 1,000 samples were collected in each bin, using an extended fluctuation of 0.02 Å, an extended time constant of 100 fs, a bias temperature of 4,000 K depositing the Gaussian biasing potentials with an initial height of 0.1 kcal/mol, a width of 0.2 Å and a frequency of 1 every 1,000 steps. For the following iterations, we performed 10 ns PCV simulations for each CCS. The PCV was sampled between 0.05 and 0.95, with a bin width of 0.01. Harmonic walls were set to restrain the PCV within 0.05 and 0.95 using force constants of 1 kcal/mol, and $\zeta$ below 500 degree[2] using a force constant of 0.001 kcal/mol degree[4]. The biasing force was applied after 1,000 samples were collected in each bin, using an extended fluctuation of 0.01, an extended time constant of 200 fs, a bias temperature of 4,000 K depositing the Gaussian biasing potentials with an initial height of 0.2 kcal/mol, a width of 0.02 and a frequency of 1 every 1,000 steps.

After the final iteration, 900 unbiased simulations of 200 ps each were performed starting from conformations sampled in the PCV simulations along the converged CCSs.

For the training of the VCN, for the initial step of the iterative process, the last 80 ns of sampling was used. For the following iterations, the last 5 ns of each simulation was used. The time-correlation function of the committor $q$ was calculated using a time-lag $\tau$ of 100 fs. The VCN consisted of 4 layers with 64 nodes each, and the trainings were performed with a maximum number of epochs of 5,000, with an early stopping set at 20 epochs.

For the initial sampling, 800 initial points were used to generate the gradient-guided pathways. For the following iterations, 2,000 initial points were used. The string pathways were then clustered using hierarchical clustering making use of a cut-off height of 0.3°. In the cases where the cut-off height was not clearly identifiable, a value of 0.3° was still employed and, afterwards, the clustered string pathways were merged using the Voronoi-cell criterion, followed by an additional hierarchical clustering analysis.

### E. Diels–Alder reaction

The Diels–Alder reaction of ethylene with vinylacetylene to form cyclohexadiene was simulated in implicit solvent (Dimethylformamide) using GFN2-xTB. Enhance sampling along the RMSDs defined with respect to the basins was not utilized due to their strong degeneracy for this system, as well as the presence of convergence issues at the self-consistent field iteration level. Therefore, The initial sampling was obtained by biasing one of the distances between the carbon atoms involved in bond formation. To ensure relevant sampling of the bond formation process and to suppress dissociative configurations of the reactants, restraints were imposed on the interatomic distances corresponding to the newly formed bonds. A harmonic potential was applied with a wall positioned at 3.5 Å, using a force constant of 100

kcal/mol Å². Additionally, to prevent sampling bias arising from chemically equivalent species differing in topology, a torsional restraint was introduced. This was defined via the dihedral angle formed by atoms 6–2–0–1 and confined within the interval of $-80°$ to $+80°$, enforced using harmonic walls with a force constant of 10 kcal/(mol degree²). The simulation employed 32 walkers, each propagated for 1 ns, with gradient sharing occurring every 1,000 steps. Langevin dynamics were used to maintain a temperature of 300 K, applying a damping coefficient of 10 ps⁻¹. For the extended Langevin formulation, parameters were set to a damping coefficient of 1 ps⁻¹, an extended fluctuation magnitude of 0.01 Å, and an extended timescale of 200 fs. The sampling was discretized into 73 bins with the width of 0.05 Å. The biasing force was applied once each bin accumulated a threshold of 5,000 samples. In the well-tempered metadynamics component, the bias temperature was set to 2,000 K, with hills deposited every 1,000 steps, having a Gaussian width of 0.028 Å, and a hill height of 0.1 kcal/mol.

For the following iterations, we performed 16 ns (0.5 ns per walker) PCV simulations along the determined CCSs connecting the two basins ($A = \{2.8 \text{ Å} \leq d_1, d_2 \leq 3.6 \text{ Å}\}$ and $B = \{1.4 \text{ Å} \leq d_1, d_2 \leq 1.8 \text{ Å}\}$). The PCV was sampled between 0.0 and 1.0, with a bin width of 0.01. Harmonic walls were set to restrain the PCV within 0.0 and 1.0 using a force constant of 1 kcal/mol. The biasing force was applied after 5,000 samples were collected in each bin, using an extended fluctuation of 0.01, an extended time constant of 200 fs, a bias temperature of 14,000 K depositing the Gaussian biasing potentials with an initial height of 0.1 kcal/mol, a width of 1.0 and a frequency of 1 every 1,000 steps.

After the final iteration, 1,000 unbiased simulations of 10 ps each were performed starting from conformations sampled in the PCV simulations along the converged CCSs.

For the training of the VCNs, the last 0.4 ns of sampling from each walker were used. The time-correlation function of the committor $q$ was calculated using a time-lag $\tau$ of 100 fs. The VCN was trained with a maximum number of epochs of 15,000 and with an early-stopping number of epochs set at 20.

For the each sampling, 500 initial points were used to generate the gradient-guided pathways. The string pathways were then clustered using hierarchical clustering using a cut-off height of 0.17 Å. The clustered string pathways were then merged according to the exchange between Voronoi cells, after which an additional hierarchical clustering analysis was performed.

## S2.  PATHWAYS IN TRIALANINE ISOMERIZATION

In Table S1, we show the main metastable states of trialanine isomerization in vacuum, which are also depicted in Fig. S1 within a reference free-energy landscape. Moreover, the 21 pathways obtained by the presented methodology are described schematically in Table S2.

TABLE S1. Positions in the ($\phi_1$, $\phi_2$, $\phi_3$) subspace of the main metastable states of trialanine in vacuum [14].

|       | $\phi_1(°)$ | $\phi_2(°)$ | $\phi_3(°)$ |
|-------|-------------|-------------|-------------|
| A     | 60          | -70         | 60          |
| B     | -70         | 60          | -70         |
| $M_1$ | -70         | -70         | 60          |
| $M_2$ | 60          | -70         | -70         |
| $M_3$ | -70         | -70         | -70         |
| $M_4$ | -70         | 60          | 60          |
| $M_5$ | 60          | 60          | 60          |
| $M_6$ | 60          | 60          | -70         |

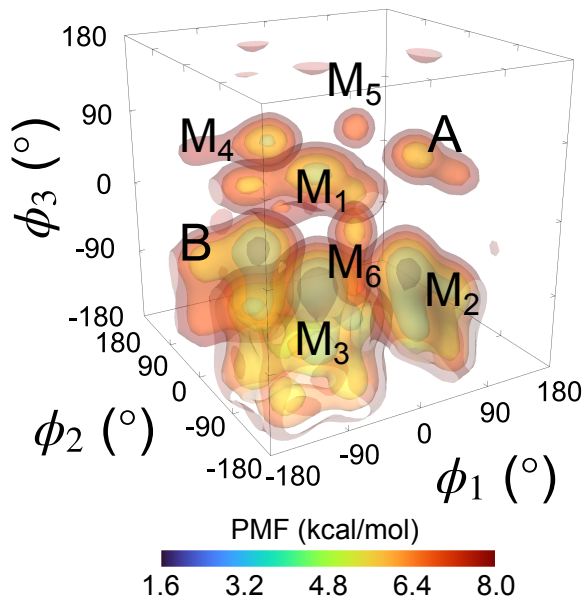

FIG. S1. Free energy landscape of trialanine in vacuum projected onto the $(\phi_1, \phi_2, \phi_3)$ subspace obtained from a WTM-eABF simulation using $\phi_1$, $\phi_2$, and $\phi_3$ as CVs. All the main metastable states (see Table S1 and Ref. [14]) are labelled.

TABLE S2. Schematic description of the converged CCSs for the isomerization of trialanine in vacuum. Each path is described through the metastable states encountered (see Table S1 and Fig. S1). "−" indicates trialanine transitions between the two basins without crossing the periodic boundaries of the CVs; otherwise, the transition between the basins is indicated with "=".

| Path label | Path description |
|:---:|:---:|
| 1 | A=$M_1$−$M_4$=B |
| 2 | A=$M_1$−$M_4$−B |
| 3 | A−$M_5$−$M_4$=B |
| 4 | A−$M_1$−$M_4$=B |
| 5 | A−$M_5$−$M_4$−B |
| 6 | A−$M_1$−$M_4$−B |
| 7 | A−$M_2$−$M_6$=B |
| 8 | A−$M_2$−$M_6$−B |
| 9 | A=$M_2$−$M_6$−B |
| 10 | A=$M_2$−$M_3$−B |
| 11 | A−$M_2$−$M_3$=B |
| 12 | A−$M_2$−$M_3$−B |
| 13 | A=$M_3$=B |
| 14 | A=$M_2$=$M_3$−B |
| 15 | A=$M_1$=$M_3$−B |
| 16 | A−$M_1$=$M_3$=B |
| 17 | A−$M_1$=$M_3$−B |
| 18 | A=$M_1$−$M_3$=B |
| 19 | A−$M_1$−$M_3$=B |
| 20 | A=$M_1$−$M_3$−B |
| 21 | A−$M_1$−$M_3$−B |

## S3. CONVERGENCE OF THE WORKFLOW

In Table S3, we present the KLD values for the different iterations of the studied process. In all of them, within 3 iterations the values constantly decrease in one or more orders of magnitude from the initial sampling, indicating a convergence of the method.

TABLE S3. Value of the KLD, $\mathcal{D}^{(k)}$, (see Eq. 3 of the main text) for the different systems and iterations.

| Iteration, $k$ | Triple well | NANMA | Trialanine | Diels–Alder |
|---|---|---|---|---|
| 1 | $3.52 \times 10^{-1}$ | $3.62 \times 10^{-1}$ | 2.35 | $4.73 \times 10^{-2}$ |
| 2 | $5.49 \times 10^{-3}$ | $7.78 \times 10^{-2}$ | 1.22 | $1.94 \times 10^{-2}$ |
| 3 | $2.58 \times 10^{-3}$ | $1.00 \times 10^{-2}$ | $8.65 \times 10^{-1}$ | $6.92 \times 10^{-4}$ |

## S4. COMBINING WTM-EABF-PCV CALCULATIONS INTO A COMPLETE EQUILIBRIUM DISTRIBUTION

### A. Histogram-less WHAM

We wish to perform enhanced sampling in a subspace of CVs, $\mathbf{z}$, using $N_w$ independent WTM-eABF-PCV calculations along different pathways. The result of the $i$–th calculation yields the biasing potentials $w_i(\mathbf{z})$. The PCV is defined from a discretized string $\{\tilde{\mathbf{z}}_k\}_{k=1}^m$ as [11],

$$s(\mathbf{z}) = \sum_{k=1}^{m} \frac{k-1}{m-1} \left( \frac{e^{-\alpha\, d(\mathbf{z}, \tilde{\mathbf{z}}_k)^2}}{\sum_{\ell=1}^{m} e^{-\alpha\, d(\mathbf{z}, \tilde{\mathbf{z}}_\ell)^2}} \right), \tag{S5}$$

where $d(\cdot, \cdot)$ is a distance, and $\alpha \in \mathbb{R}^+$ is a positive constant that controls the smoothness of the quantity. In addition, one defines an orthogonal variable,

$$\zeta(\mathbf{z}) = -\frac{1}{\alpha} \ln \sum_{\ell=1}^{m} e^{-\alpha\, d(\mathbf{z}, \tilde{\mathbf{z}}_\ell)^2}. \tag{S6}$$

Hence, for each string $i$, one can define a $s_i(\mathbf{z})$ and $\zeta_i(\mathbf{z})$, from which the $i$-th PCV calculation is carried out. Once it is converged, the ABF calculation flatten the free energy landscape along the $s_i$ coordinate (cancels the PMF of the system along $s_i$) while the biasing potential orthogonal to the reaction tube remains. Therefore, the converged biasing potential $w_i(\mathbf{z})$ is given by,

$$w_i(\mathbf{z}) = -V_{\mathrm{b}}[s_i(\mathbf{z})] + u_\perp[\zeta_i(\mathbf{z})], \tag{S7}$$

where $V_{\mathrm{b}}$ is the one-dimensional biasing potential determined by the enhanced sampling method (WTM-eABF in the particular case of this work), and $u_\perp$ refers to the direction perpendicular to the direction of application of $V_{\mathrm{b}}$.

The individual unbiased equilibrium distribution function in the space $\mathbf{z}$ coincides with,

$$\langle \rho(\mathbf{z}) \rangle_{\mathrm{unb}}^{(i)} = e^{+\beta w_i(\mathbf{z})} \langle \rho(\mathbf{z}) \rangle_{\mathrm{b}}^{(i)} e^{-\beta F_i}, \tag{S8}$$

where $\langle \rho(\mathbf{z}) \rangle^{(i)}$ is the $i$–th biased distribution and $F_i$ is some constant to be determined.

The WHAM equations express the optimal estimate for the unbiased distribution function as a $\mathbf{z}$-dependent weighted sum over the $N_w$ individual unbiased distribution functions:

$$\begin{aligned}
\langle \rho(\mathbf{z}) \rangle &= \sum_{i=1}^{N_w} \langle \rho(\mathbf{z}) \rangle_{\mathrm{unb}}^{(i)} \left( \frac{n_i e^{-\beta(w_i(\mathbf{z}) - F_i)}}{\sum_{j=1}^{N_w} n_j e^{-\beta(w_j(\mathbf{z}) - F_j)}} \right) \\
&= \sum_{i=1}^{N_w} n_i \langle \rho(\mathbf{z}) \rangle_{\mathrm{b}}^{(i)} \left( \sum_{j=1}^{N_w} n_j e^{-\beta(w_j(\mathbf{z}) - F_j)} \right)^{-1} \\
&= H_{\mathrm{b}}(\mathbf{z}) R(\mathbf{z}),
\end{aligned} \tag{S9}$$

where $n_i$ is the number of independent data points used to construct the biased distribution function, $H_{\mathrm{b}}(\mathbf{z})$ is the cumulative histogram of all the biased simulations,

$$H_{\mathrm{b}}(\mathbf{z}) = \sum_{i=1}^{N_w} n_i \langle \rho(\mathbf{z}) \rangle_{\mathrm{b}}^{(i)} \rightarrow \sum_{t} \delta(\mathbf{z}_{\mathrm{b}}(t) - \mathbf{z}), \tag{S10}$$

where $\mathbf{z}_b(t)$ refers to the value, in CV-space, of the biased trajectory at time $t$, and $R(\mathbf{z})$ is the reweighing factor,

$$R(\mathbf{z}) = \left( \sum_{j=1}^{N_w} n_j e^{-\beta(w_j(\mathbf{z}) - F_j)} \right)^{-1}. \tag{S11}$$

Then, in effect, we have

$$\langle \rho(\mathbf{z}) \rangle \rightarrow \sum_t \delta(\mathbf{z}_b(t) - \mathbf{z}) R(\mathbf{z}_b(t)), \tag{S12}$$

where $t$ are the snapshots along the WTM-eABF-PCV trajectories. The free energy constants $F_i$, needed in Eq. (S9), are determined using the optimal estimate for the distribution function:

$$e^{-\beta F_i} = \int d\mathbf{z} \, e^{-\beta w_i(\mathbf{z})} \langle \rho(\mathbf{z}) \rangle. \tag{S13}$$

Since the distribution function itself depends on the set of constants $\{F_j\}$, the WHAM Eqs. (S9) and (S13) must be solved self-consistently through an iteration procedure. In practice we could set one of the $F_i$ to a fixed value. Let us assume that we set $F_1 = 0$, then the free energy constants $F_i$, needed in Eq. (2), are determined using,

$$e^{-\beta(F_i - F_1)} = \frac{\int d\mathbf{z} \, e^{-\beta w_i(\mathbf{z})} \langle \rho(\mathbf{z}) \rangle}{\int d\mathbf{z} \, e^{-\beta w_1(\mathbf{z})} \langle \rho(\mathbf{z}) \rangle} \approx \frac{\sum_t e^{-\beta w_i(\mathbf{z}_b(t))} R(\mathbf{z}_b(t))}{\sum_{t'} e^{-\beta w_1(\mathbf{z}_b(t'))} R(\mathbf{z}_b(t'))}. \tag{S14}$$

This is the so-called histogram-less version of WHAM. Starting from an initial guess for the $N_w$ free energy constants $F_i$, an estimate for the unbiased distribution is obtained from Eq. (S9). This estimate for $\langle \rho(\mathbf{z}) \rangle$ is used in Eq. (S13) to generate new estimates for the $N_w$ free energy constants $F_i$, and a new unbiased distribution is generated with Eq. (S9). The iteration cycle is repeated until both equations are satisfied.

### 1. Time correlation function

Once the unbiased equilibrium distribution has been obtained, it is possible to calculate the unbiased time-correlation function of the committor, $C[q; t]$,

$$C[q; t] = \frac{1}{2} \left\langle (q(t) - q(0))^2 \right\rangle. \tag{S15}$$

We express the time-correlation function of interest as an average over equilibrium initial conditions $\mathbf{z}_i$, picked from the configurations along the biased ABF trajectories $\mathbf{z}_b(t)$:

$$C[q; t] = \int d\mathbf{z} \, \langle \rho(\mathbf{z}) \rangle \, c[q(\mathbf{z}_i); t] \approx \frac{\sum_i R(\mathbf{z}_i) \, c_{\mathbf{z}_i}[q; t]}{\sum_j R(\mathbf{z}_j)}, \quad c[q(\mathbf{z}_i); t] \equiv c_{\mathbf{z}_i}[q; t] = \frac{1}{n_t} \sum_{k=0}^{n_t} \left( q(\mathbf{z}(t + k\Delta t | \mathbf{z}_i)) - q(\mathbf{z}(k\Delta t | \mathbf{z}_i)) \right)^2, \tag{S16}$$

where the time-correlation function $c_{\mathbf{z}_i}[q; t]$ is calculated as an average over time from a new unbiased trajectory initiated at $\mathbf{z}_i$, where $\Delta t$ represents the time step of the trajectories. Finally, once this is done and we learned a committor $q$, and we have $C(t) \equiv C[q; t]$, the steady-state flux can be recovered as

$$J_{AB} = \lim_{t \to \tau_q} \dot{C}(t), \tag{S17}$$

with $\tau_q$ a finite relaxation time, not necessary the same as the time lag used in the VCN.

### B. Results

### S5. ESTIMATION OF THE RATE CONSTANTS

To estimate the rate constant for the transition process at hand, the discussion in Sec. S4 A 1 is used. After a final set of unbiased trajectories initiated from randomly selected initial conditions from the PCV calculations in

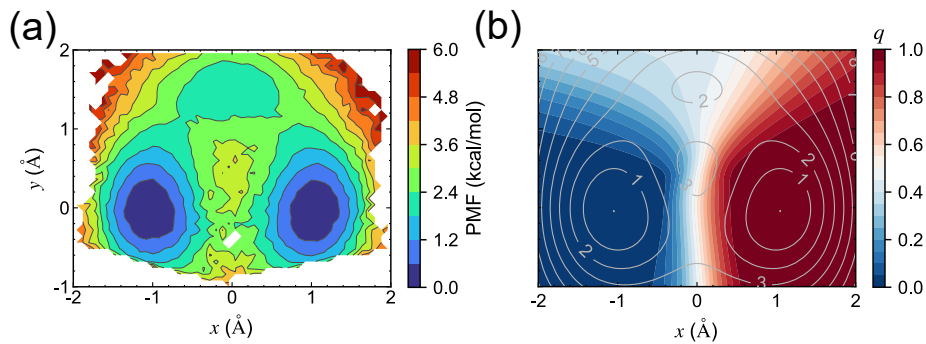

FIG. S2. (a) Triple-well potential obtained from bin-less WHAM combining the sampling of the PCV simulations along the converged CCSs. (b) Learned committor, $q$ ,for the triple-well potential obtained from short unbiased simulations by shooting from points sampled during the PCV simulations along the converged CCSs. The gray lines in (b) correspond to different contour lines of the triple-well potential.

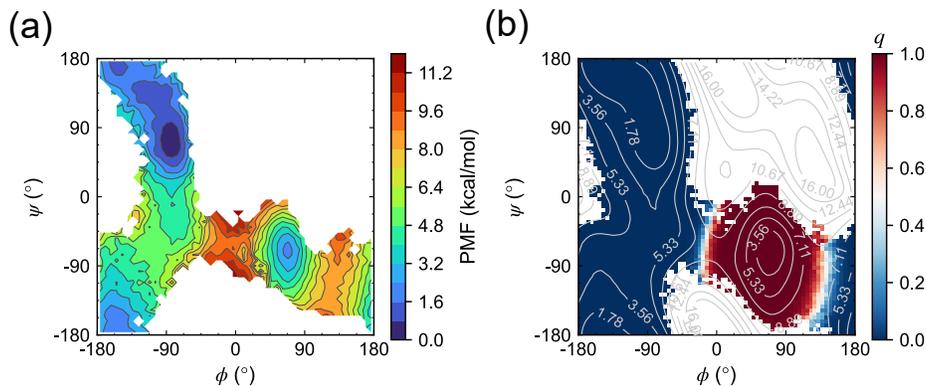

FIG. S3. (a) Free energy landscape and (b) learned committor, $q$, of NANMA isomerization in vacuum projected onto the $(\phi, \psi)$-subspace and with $\theta = \omega = 0°$ obtained from the histogram-less WHAM combining the sampling of the different PCV calculations along the converged CCSs. The gray lines in (b) correspond to different contour lines of a reference free energy landscape.

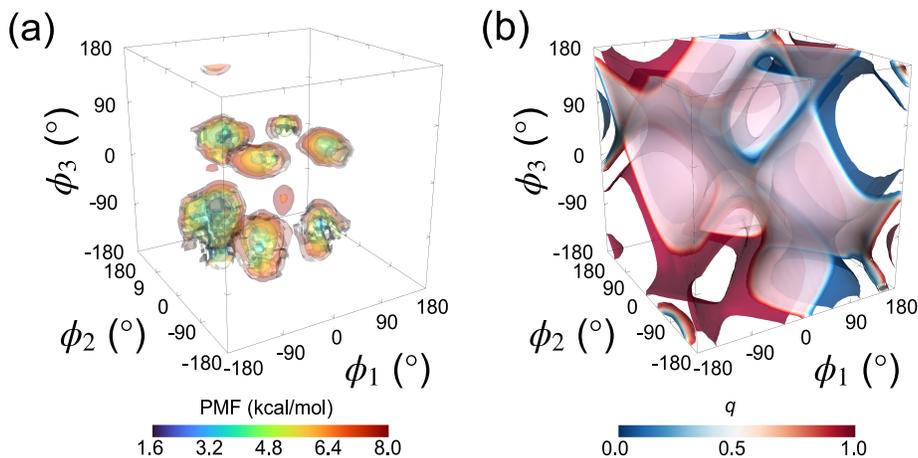

FIG. S4. (a) Free energy landscape of trialanine in vacuum projected onto the $(\phi_1, \phi_2, \phi_3)$-subspace obtained from bin-less WHAM combining the sampling of the PCV simulations along the converged CCSs. (b) Learned committor $q$ for the $A \rightarrow B$ isomerization of trialanine in vacuum projected onto the $(\phi_1, \phi_2, \phi_3)$-subspace obtained from short unbiased simulations by shooting from points sampled during the PCV simulations along the converged CCSs.

the last iteration of the workflow, we learned a committor using the VCN, and compute $p_A$ and $C[q, \tau]$ for different values of $\tau$. In Figs. S2–S4, we show the free-energy landscapes obtained by using the WHAM, as well as the final learned committor from the unbiased trajectories simulated after the convergence of the methodology for the triple-well potential and NANMA and trialanine isomerizations, respectively. In the case of the Diels–Alder reaction, there is no need of usage of WHAM, since a single string pathway is obtained. The estimates are computed following Eq. S16. Moreover, the variance of the correlation function is defined as follows [15],

$$\mathrm{Var}[C[q, \tau]] = \frac{\sum_i R^2(\mathbf{z}_i)\mathrm{Var}\left[c_{\mathbf{z}_i}[q, \tau]\right]}{\left(\sum_j R(\mathbf{z}_j)\right)^2}. \tag{S18}$$

In every particular computation, $\mathrm{Var}\left[c_{\mathbf{z}_i}[q, \tau]\right]$ is estimated with the corresponding sampling variance for the $i$-th unbiased trajectory. Then, the error associated with each punctual estimate of the total correlation function is identified with the corresponding standard deviation, i.e., the square root of the estimated variance.

In order to ensure the linear regime in which we can compute the slope in Eq. (S17), we identify the end of a first relaxation process related to the equilibration of the initial structures. Once these initial relaxation times are appropriately selected, we compute the slope and its uncertainty using bootstrapping [16, 17] on the estimator of the associated linear regression. The reason of using a bootstrapping resampling is motivated by the presence of heterocedasticity. The bootstrap is performed using 10,000 resamplings on the selected points for the linear regression. The estimate of the slope is identified with the confidence interval at 99% of confidence level. Due to the assumption of Gaussian distribution for each point used in the linear regression, the intervals are symmetric around the sampling mean. Moreover, it is worth noting that, in the case of the two-dimensional triple-well potential, there is no need to wait for a first relaxation due to the absence of molecular structure. In Fig. S5, we show the results of the rate constants estimates for the different transitions.

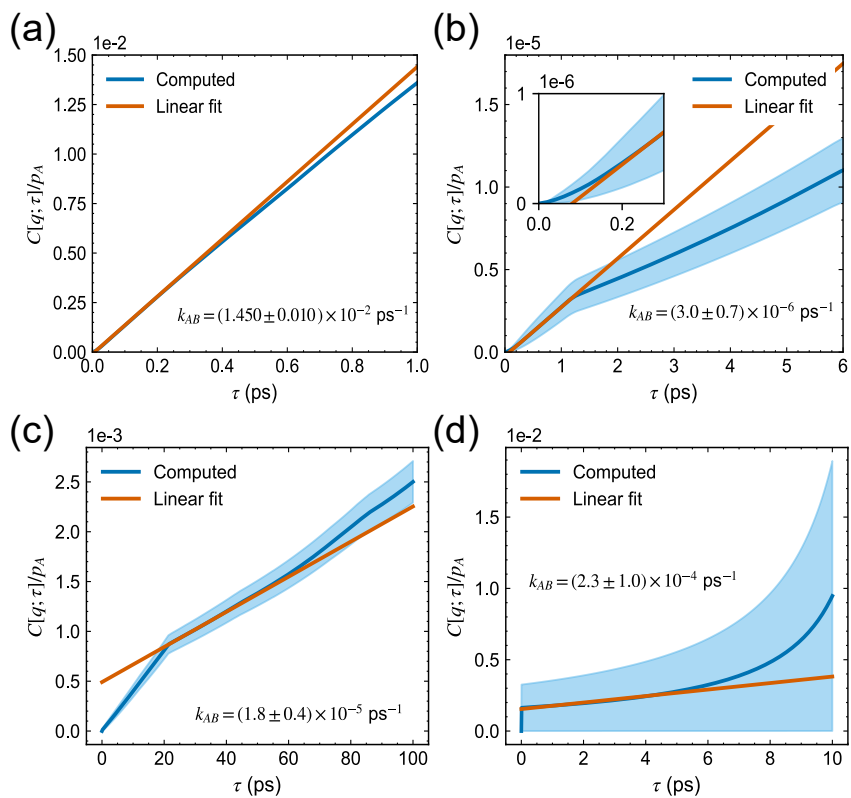

FIG. S5. Rate constant $k_{AB}$ determined from the resulting committors learned by the VCN trained on short unbiased simulations of: (a) the triple-well potential; (b) the isomerization of NANMA; (c) trialanine conformational equilibrium in vacuum; (d) Diels–Alder reaction of vinyl-acetylene and butyne in implicit solvent (dimethylformamide).

[1] J. C. Phillips, D. J. Hardy, J. D. C. Maia, J. E. Stone, J. V. Ribeiro, R. C. Bernardi, R. Buch, G. Fiorin, J. Hénin, W. Jiang, R. McGreevy, M. C. R. Melo, B. K. Radak, R. D. Skeel, A. Singharoy, Y. Wang, B. Roux, A. Aksimentiev, Z. Luthey-Schulten, L. V. Kalé, K. Schulten, C. Chipot, and E. Tajkhorshid, Scalable molecular dynamics on cpu and gpu architectures with NAMD, J. Chem. Phys. **153**, 044130 (2020).

[2] G. Fiorin, F. Marinelli, L. R. Forrest, H. Chen, C. Chipot, A. Kohlmeyer, H. Santuz, and J. Hénin, Expanded functionality and portability for the colvars library, J. Phys. Chem. B **128**, 11108 (2024).

[3] H. Fu, X. Shao, W. Cai, and C. Chipot, Taming rugged free energy landscapes using an average force, Acc. Chem. Res. **52**, 3254 (2019).

[4] K. Minoukadeh, C. Chipot, and T. Lelièvre, Potential of mean force calculations: A multiple-walker adaptive biasing force approach, J. Chem. Theory Comput. **6**, 1008 (2010).

[5] H. C. Andersen, RATTLE: A "velocity" version of the shake algorithm for molecular dynamics calculations, J. Comput. Phys. **52**, 24 (1983).

[6] H. Chen, B. Roux, and C. Chipot, Discovering reaction pathways, slow variables, and committor probabilities with machine learning, J. Chem. Theory Comput. **19**, 4414 (2023).

[7] D. Clevert, T. Unterthiner, and S. Hochreiter, Fast and accurate deep network learning by exponential linear units (elus), in *4th International Conference on Learning Representations, ICLR 2016, San Juan, Puerto Rico, May 2-4, 2016, Conference Track Proceedings*, edited by Y. Bengio and Y. LeCun (2016).

[8] P. V. Banushkina and S. V. Krivov, Optimal reaction coordinates, Wiley Interdiscip. Rev. Comput. Mol. Sci. **6**, 748 (2016).

[9] B. Roux, Transition rate theory, spectral analysis, and reactive paths, J. Chem. Phys **156**, 134111 (2022).

[10] A. Megías, S. Contreras Arredondo, C. G. Chen, C. Tang, B. Roux, and C. Chipot, Iterative variational learning of committor-consistent transition pathways using artificial neural networks, Nat. Comput. Sci. **5**, 592–602 (2025).

[11] D. Branduardi, F. L. Gervasio, and M. Parrinello, From A to B in free energy space, J. Chem. Phys. **126**, 054103 (2007).

[12] A. D. J. MacKerell, D. Bashford, M. Bellott, R. L. J. Dunbrack, J. D. Evanseck, M. J. Field, S. Fischer, J. Gao, H. Guo, S. Ha, D. Joseph-McCarthy, L. Kuchnir, K. Kuczera, F. T. K. Lau, C. Mattos, S. Michnick, T. Ngo, D. T. Nguyen, B. Prodhom, W. E. Reiher, B. Roux, M. Schlenkrich, J. C. Smith, R. Stote, J. Straub, M. Watanabe, J. Wiórkiewicz-Kuczera, D. Yin, and M. Karplus, All-atom empirical potential for molecular modeling and dynamics studies of proteins, J. Phys. Chem. B **102**, 3586 (1998).

[13] J. A. Maier, C. Martinez, K. Kasavajhala, L. Wickstrom, K. E. Hauser, and C. Simmerling, ff14sb: Improving the accuracy of protein side chain and backbone parameters from ff99sb, J. Chem. Theory Comput. **11**, 3696 (2015).

[14] H. Chen, D. Ogden, S. Pant, W. Cai, E. Tajkhorshid, M. Moradi, B. Roux, and C. Chipot, A companion guide to the string method with swarms of trajectories: Characterization, performance, and pitfalls, J. Chem. Theory Comput. **18**, 1406 (2022).

[15] J. Harris and B. Roux, Membrane permeability of sucrose calculated from equilibrium time-correlation functions using molecular dynamics simulations with enhanced sampling, J. Phys. Chem. B **129**, 7172 (2025).

[16] B. Efron, Bootstrap methods: Another look at the jackknife, Ann. Statist. **7**, 1 (1979).

[17] P. Young, *Everything You Wanted to Know About Data Analysis and Fitting but Were Afraid to Ask*, SpringerBriefs in Physics (Springer, Cham, 2015).




\begin{thebibliography}{40}%
\makeatletter
\providecommand \@ifxundefined [1]{%
 \@ifx{#1\undefined}
}%
\providecommand \@ifnum [1]{%
 \ifnum #1\expandafter \@firstoftwo
 \else \expandafter \@secondoftwo
 \fi
}%
\providecommand \@ifx [1]{%
 \ifx #1\expandafter \@firstoftwo
 \else \expandafter \@secondoftwo
 \fi
}%
\providecommand \natexlab [1]{#1}%
\providecommand \enquote  [1]{``#1''}%
\providecommand \bibnamefont  [1]{#1}%
\providecommand \bibfnamefont [1]{#1}%
\providecommand \citenamefont [1]{#1}%
\providecommand \href@noop [0]{\@secondoftwo}%
\providecommand \href [0]{\begingroup \@sanitize@url \@href}%
\providecommand \@href[1]{\@@startlink{#1}\@@href}%
\providecommand \@@href[1]{\endgroup#1\@@endlink}%
\providecommand \@sanitize@url [0]{\catcode `\\12\catcode `\$12\catcode `\&12\catcode `\#12\catcode `\^12\catcode `\_12\catcode `\%12\relax}%
\providecommand \@@startlink[1]{}%
\providecommand \@@endlink[0]{}%
\providecommand \url  [0]{\begingroup\@sanitize@url \@url }%
\providecommand \@url [1]{\endgroup\@href {#1}{\urlprefix }}%
\providecommand \urlprefix  [0]{URL }%
\providecommand \Eprint [0]{\href }%
\providecommand \doibase [0]{https://doi.org/}%
\providecommand \selectlanguage [0]{\@gobble}%
\providecommand \bibinfo  [0]{\@secondoftwo}%
\providecommand \bibfield  [0]{\@secondoftwo}%
\providecommand \translation [1]{[#1]}%
\providecommand \BibitemOpen [0]{}%
\providecommand \bibitemStop [0]{}%
\providecommand \bibitemNoStop [0]{.\EOS\space}%
\providecommand \EOS [0]{\spacefactor3000\relax}%
\providecommand \BibitemShut  [1]{\csname bibitem#1\endcsname}%
\let\auto@bib@innerbib\@empty
\bibitem [{\citenamefont {Moradi}\ and\ \citenamefont {Tajkhorshid}(2013)}]{moradi2013}%
  \BibitemOpen
  \bibfield  {author} {\bibinfo {author} {\bibfnamefont {M.}~\bibnamefont {Moradi}}\ and\ \bibinfo {author} {\bibfnamefont {E.}~\bibnamefont {Tajkhorshid}},\ }\bibfield  {title} {\bibinfo {title} {Mechanistic picture for conformational transition of a membrane transporter at atomic resolution},\ }\href {https://doi.org/10.1073/pnas.1313202110} {\bibfield  {journal} {\bibinfo  {journal} {Proc. Natl. Acad. Sci. U.S.A.}\ }\textbf {\bibinfo {volume} {110}},\ \bibinfo {pages} {18916} (\bibinfo {year} {2013})}\BibitemShut {NoStop}%
\bibitem [{\citenamefont {Lindorff-Larsen}\ \emph {et~al.}(2011)\citenamefont {Lindorff-Larsen}, \citenamefont {Piana}, \citenamefont {Dror},\ and\ \citenamefont {Shaw}}]{Lindorff-Larsen2011fold}%
  \BibitemOpen
  \bibfield  {author} {\bibinfo {author} {\bibfnamefont {K.}~\bibnamefont {Lindorff-Larsen}}, \bibinfo {author} {\bibfnamefont {S.}~\bibnamefont {Piana}}, \bibinfo {author} {\bibfnamefont {R.~O.}\ \bibnamefont {Dror}},\ and\ \bibinfo {author} {\bibfnamefont {D.~E.}\ \bibnamefont {Shaw}},\ }\bibfield  {title} {\bibinfo {title} {{How Fast-Folding Proteins Fold}},\ }\href {https://doi.org/10.1126/science.1208351} {\bibfield  {journal} {\bibinfo  {journal} {Science}\ }\textbf {\bibinfo {volume} {334}},\ \bibinfo {pages} {517} (\bibinfo {year} {2011})}\BibitemShut {NoStop}%
\bibitem [{\citenamefont {Ajaz}\ \emph {et~al.}(2011)\citenamefont {Ajaz}, \citenamefont {Bradley}, \citenamefont {Burrell}, \citenamefont {Li}, \citenamefont {Daoust}, \citenamefont {Bovee}, \citenamefont {DiRico},\ and\ \citenamefont {Johnson}}]{ajaz_2011}%
  \BibitemOpen
  \bibfield  {author} {\bibinfo {author} {\bibfnamefont {A.}~\bibnamefont {Ajaz}}, \bibinfo {author} {\bibfnamefont {A.~Z.}\ \bibnamefont {Bradley}}, \bibinfo {author} {\bibfnamefont {R.~C.}\ \bibnamefont {Burrell}}, \bibinfo {author} {\bibfnamefont {W.~H.~H.}\ \bibnamefont {Li}}, \bibinfo {author} {\bibfnamefont {K.~J.}\ \bibnamefont {Daoust}}, \bibinfo {author} {\bibfnamefont {L.~B.}\ \bibnamefont {Bovee}}, \bibinfo {author} {\bibfnamefont {K.~J.}\ \bibnamefont {DiRico}},\ and\ \bibinfo {author} {\bibfnamefont {R.~P.}\ \bibnamefont {Johnson}},\ }\bibfield  {title} {\bibinfo {title} {Concerted vs stepwise mechanisms in dehydro-diels–alder reactions},\ }\href {https://doi.org/10.1021/jo201567d} {\bibfield  {journal} {\bibinfo  {journal} {J. Org. Chem.}\ }\textbf {\bibinfo {volume} {76}},\ \bibinfo {pages} {9320} (\bibinfo {year} {2011})}\BibitemShut {NoStop}%
\bibitem [{\citenamefont {Yang}\ \emph {et~al.}(2009)\citenamefont {Yang}, \citenamefont {Banavali},\ and\ \citenamefont {Roux}}]{yang2009}%
  \BibitemOpen
  \bibfield  {author} {\bibinfo {author} {\bibfnamefont {S.}~\bibnamefont {Yang}}, \bibinfo {author} {\bibfnamefont {N.~K.}\ \bibnamefont {Banavali}},\ and\ \bibinfo {author} {\bibfnamefont {B.}~\bibnamefont {Roux}},\ }\bibfield  {title} {\bibinfo {title} {Mapping the conformational transition in src activation by cumulating the information from multiple molecular dynamics trajectories},\ }\href {https://doi.org/10.1073/pnas.0808261106} {\bibfield  {journal} {\bibinfo  {journal} {Proc. Natl. Acad. Sci. U.S.A.}\ }\textbf {\bibinfo {volume} {106}},\ \bibinfo {pages} {3776} (\bibinfo {year} {2009})}\BibitemShut {NoStop}%
\bibitem [{\citenamefont {Wulferding}\ \emph {et~al.}(2017)\citenamefont {Wulferding}, \citenamefont {Kim}, \citenamefont {Yang}, \citenamefont {Jeong}, \citenamefont {Barros}, \citenamefont {Kato}, \citenamefont {Martin}, \citenamefont {Ayala-Valenzuela}, \citenamefont {Lee}, \citenamefont {Choi}, \citenamefont {Ronning}, \citenamefont {Civale}, \citenamefont {Baumbach}, \citenamefont {Bauer}, \citenamefont {Thompson}, \citenamefont {Movshovich},\ and\ \citenamefont {Kim}}]{Wulferding2017}%
  \BibitemOpen
  \bibfield  {author} {\bibinfo {author} {\bibfnamefont {D.}~\bibnamefont {Wulferding}}, \bibinfo {author} {\bibfnamefont {H.}~\bibnamefont {Kim}}, \bibinfo {author} {\bibfnamefont {I.}~\bibnamefont {Yang}}, \bibinfo {author} {\bibfnamefont {J.}~\bibnamefont {Jeong}}, \bibinfo {author} {\bibfnamefont {K.}~\bibnamefont {Barros}}, \bibinfo {author} {\bibfnamefont {Y.}~\bibnamefont {Kato}}, \bibinfo {author} {\bibfnamefont {I.}~\bibnamefont {Martin}}, \bibinfo {author} {\bibfnamefont {O.~E.}\ \bibnamefont {Ayala-Valenzuela}}, \bibinfo {author} {\bibfnamefont {M.}~\bibnamefont {Lee}}, \bibinfo {author} {\bibfnamefont {H.~C.}\ \bibnamefont {Choi}}, \bibinfo {author} {\bibfnamefont {F.}~\bibnamefont {Ronning}}, \bibinfo {author} {\bibfnamefont {L.}~\bibnamefont {Civale}}, \bibinfo {author} {\bibfnamefont {R.~E.}\ \bibnamefont {Baumbach}}, \bibinfo {author} {\bibfnamefont {E.~D.}\ \bibnamefont {Bauer}}, \bibinfo {author} {\bibfnamefont {J.~D.}\ \bibnamefont {Thompson}}, \bibinfo {author} {\bibfnamefont
  {R.}~\bibnamefont {Movshovich}},\ and\ \bibinfo {author} {\bibfnamefont {J.}~\bibnamefont {Kim}},\ }\bibfield  {title} {\bibinfo {title} {Domain engineering of the metastable domains in the 4f-uniaxial-ferromagnet {CeRu2Ga2B}},\ }\href {https://doi.org/10.1038/srep46296} {\bibfield  {journal} {\bibinfo  {journal} {Sci. Rep.}\ }\textbf {\bibinfo {volume} {7}},\ \bibinfo {pages} {46296} (\bibinfo {year} {2017})}\BibitemShut {NoStop}%
\bibitem [{\citenamefont {Biroli}\ and\ \citenamefont {Kurchan}(2001)}]{BK01}%
  \BibitemOpen
  \bibfield  {author} {\bibinfo {author} {\bibfnamefont {G.}~\bibnamefont {Biroli}}\ and\ \bibinfo {author} {\bibfnamefont {J.}~\bibnamefont {Kurchan}},\ }\bibfield  {title} {\bibinfo {title} {Metastable states in glassy systems},\ }\href {https://doi.org/10.1103/PhysRevE.64.016101} {\bibfield  {journal} {\bibinfo  {journal} {Phys. Rev. E}\ }\textbf {\bibinfo {volume} {64}},\ \bibinfo {pages} {016101} (\bibinfo {year} {2001})}\BibitemShut {NoStop}%
\bibitem [{\citenamefont {Lucente}\ \emph {et~al.}(2022)\citenamefont {Lucente}, \citenamefont {Herbert},\ and\ \citenamefont {Bouchet}}]{lucente2022climate}%
  \BibitemOpen
  \bibfield  {author} {\bibinfo {author} {\bibfnamefont {D.}~\bibnamefont {Lucente}}, \bibinfo {author} {\bibfnamefont {C.}~\bibnamefont {Herbert}},\ and\ \bibinfo {author} {\bibfnamefont {F.}~\bibnamefont {Bouchet}},\ }\bibfield  {title} {\bibinfo {title} {{Committor functions for climate phenomena at the predictability margin: The example of El Ni{\~n}o--Southern Oscillation in the Jin and Timmermann model}},\ }\href {https://doi.org/10.1175/JAS-D-22-0038.1} {\bibfield  {journal} {\bibinfo  {journal} {J. Atmos. Sci}\ }\textbf {\bibinfo {volume} {79}},\ \bibinfo {pages} {2387} (\bibinfo {year} {2022})}\BibitemShut {NoStop}%
\bibitem [{\citenamefont {Chen}\ and\ \citenamefont {Chipot}(2022)}]{chen_enhancing_2022}%
  \BibitemOpen
  \bibfield  {author} {\bibinfo {author} {\bibfnamefont {H.}~\bibnamefont {Chen}}\ and\ \bibinfo {author} {\bibfnamefont {C.}~\bibnamefont {Chipot}},\ }\bibfield  {title} {\bibinfo {title} {Enhancing sampling with free-energy calculations},\ }\href {https://doi.org/10.1016/j.sbi.2022.102497} {\bibfield  {journal} {\bibinfo  {journal} {Curr. Opin. Struct. Biol.}\ }\textbf {\bibinfo {volume} {77}},\ \bibinfo {pages} {102497} (\bibinfo {year} {2022})}\BibitemShut {NoStop}%
\bibitem [{\citenamefont {Rogal}(2021)}]{rogal_reaction_2021}%
  \BibitemOpen
  \bibfield  {author} {\bibinfo {author} {\bibfnamefont {J.}~\bibnamefont {Rogal}},\ }\bibfield  {title} {\bibinfo {title} {Reaction coordinates in complex systems-a perspective},\ }\href {https://doi.org/10.1140/epjb/s10051-021-00233-5} {\bibfield  {journal} {\bibinfo  {journal} {Eur. Phys. J. B}\ }\textbf {\bibinfo {volume} {94}},\ \bibinfo {pages} {223} (\bibinfo {year} {2021})}\BibitemShut {NoStop}%
\bibitem [{\citenamefont {Branduardi}\ \emph {et~al.}(2007)\citenamefont {Branduardi}, \citenamefont {Gervasio},\ and\ \citenamefont {Parrinello}}]{branduardi_b_2007}%
  \BibitemOpen
  \bibfield  {author} {\bibinfo {author} {\bibfnamefont {D.}~\bibnamefont {Branduardi}}, \bibinfo {author} {\bibfnamefont {F.~L.}\ \bibnamefont {Gervasio}},\ and\ \bibinfo {author} {\bibfnamefont {M.}~\bibnamefont {Parrinello}},\ }\bibfield  {title} {\bibinfo {title} {From {A} to {B} in free energy space},\ }\href {https://doi.org/10.1063/1.2432340} {\bibfield  {journal} {\bibinfo  {journal} {J. Chem. Phys.}\ }\textbf {\bibinfo {volume} {126}},\ \bibinfo {pages} {054103} (\bibinfo {year} {2007})}\BibitemShut {NoStop}%
\bibitem [{\citenamefont {E}\ and\ \citenamefont {Vanden-Eijnden}(2010)}]{Vanden-Eijnden-2010}%
  \BibitemOpen
  \bibfield  {author} {\bibinfo {author} {\bibfnamefont {W.}~\bibnamefont {E}}\ and\ \bibinfo {author} {\bibfnamefont {E.}~\bibnamefont {Vanden-Eijnden}},\ }\bibfield  {title} {\bibinfo {title} {{{T}ransition-path theory and path-finding algorithms for the study of rare events}},\ }\href {https://doi.org/10.1146/annurev.physchem.040808.090412} {\bibfield  {journal} {\bibinfo  {journal} {Annu. Rev. Phys. Chem.}\ }\textbf {\bibinfo {volume} {61}},\ \bibinfo {pages} {391} (\bibinfo {year} {2010})}\BibitemShut {NoStop}%
\bibitem [{\citenamefont {E}\ \emph {et~al.}(2005)\citenamefont {E}, \citenamefont {Ren},\ and\ \citenamefont {Vanden-Eijnden}}]{e_finite_2005}%
  \BibitemOpen
  \bibfield  {author} {\bibinfo {author} {\bibfnamefont {W.}~\bibnamefont {E}}, \bibinfo {author} {\bibfnamefont {W.}~\bibnamefont {Ren}},\ and\ \bibinfo {author} {\bibfnamefont {E.}~\bibnamefont {Vanden-Eijnden}},\ }\bibfield  {title} {\bibinfo {title} {Finite temperature string method for the study of rare events},\ }\href {https://doi.org/10.1021/jp0455430} {\bibfield  {journal} {\bibinfo  {journal} {J. Phys. Chem. B}\ }\textbf {\bibinfo {volume} {109}},\ \bibinfo {pages} {6688} (\bibinfo {year} {2005})}\BibitemShut {NoStop}%
\bibitem [{\citenamefont {Maragliano}\ \emph {et~al.}(2006)\citenamefont {Maragliano}, \citenamefont {Fischer}, \citenamefont {Vanden-Eijnden},\ and\ \citenamefont {Ciccotti}}]{maragliano2006string}%
  \BibitemOpen
  \bibfield  {author} {\bibinfo {author} {\bibfnamefont {L.}~\bibnamefont {Maragliano}}, \bibinfo {author} {\bibfnamefont {A.}~\bibnamefont {Fischer}}, \bibinfo {author} {\bibfnamefont {E.}~\bibnamefont {Vanden-Eijnden}},\ and\ \bibinfo {author} {\bibfnamefont {G.}~\bibnamefont {Ciccotti}},\ }\bibfield  {title} {\bibinfo {title} {{String method in collective variables: Minimum free energy paths and isocommittor surfaces}},\ }\href {https://doi.org/10.1063/1.2212942} {\bibfield  {journal} {\bibinfo  {journal} {J. Chem. Phys.}\ }\textbf {\bibinfo {volume} {125}},\ \bibinfo {pages} {024106} (\bibinfo {year} {2006})}\BibitemShut {NoStop}%
\bibitem [{\citenamefont {Pan}\ \emph {et~al.}(2008)\citenamefont {Pan}, \citenamefont {Sezer},\ and\ \citenamefont {Roux}}]{pan_finding_2008}%
  \BibitemOpen
  \bibfield  {author} {\bibinfo {author} {\bibfnamefont {A.~C.}\ \bibnamefont {Pan}}, \bibinfo {author} {\bibfnamefont {D.}~\bibnamefont {Sezer}},\ and\ \bibinfo {author} {\bibfnamefont {B.}~\bibnamefont {Roux}},\ }\bibfield  {title} {\bibinfo {title} {Finding transition pathways using the string method with swarms of trajectories},\ }\href {https://doi.org/10.1021/jp0777059} {\bibfield  {journal} {\bibinfo  {journal} {J. Phys. Chem. B}\ }\textbf {\bibinfo {volume} {112}},\ \bibinfo {pages} {3432} (\bibinfo {year} {2008})}\BibitemShut {NoStop}%
\bibitem [{\citenamefont {Fu}\ \emph {et~al.}(2020)\citenamefont {Fu}, \citenamefont {Chen}, \citenamefont {Wang}, \citenamefont {Chai}, \citenamefont {Shao}, \citenamefont {Cai},\ and\ \citenamefont {Chipot}}]{fu2020finding}%
  \BibitemOpen
  \bibfield  {author} {\bibinfo {author} {\bibfnamefont {H.}~\bibnamefont {Fu}}, \bibinfo {author} {\bibfnamefont {H.}~\bibnamefont {Chen}}, \bibinfo {author} {\bibfnamefont {X.}~\bibnamefont {Wang}}, \bibinfo {author} {\bibfnamefont {H.}~\bibnamefont {Chai}}, \bibinfo {author} {\bibfnamefont {X.}~\bibnamefont {Shao}}, \bibinfo {author} {\bibfnamefont {W.}~\bibnamefont {Cai}},\ and\ \bibinfo {author} {\bibfnamefont {C.}~\bibnamefont {Chipot}},\ }\bibfield  {title} {\bibinfo {title} {Finding an optimal pathway on a multidimensional free-energy landscape},\ }\href {https://doi.org/10.1021/acs.jcim.0cf00279} {\bibfield  {journal} {\bibinfo  {journal} {J. Chem. Inf. Model.}\ }\textbf {\bibinfo {volume} {60}},\ \bibinfo {pages} {5366} (\bibinfo {year} {2020})}\BibitemShut {NoStop}%
\bibitem [{\citenamefont {Vanden-Eijnden}\ and\ \citenamefont {Tal}(2005)}]{VT05}%
  \BibitemOpen
  \bibfield  {author} {\bibinfo {author} {\bibfnamefont {E.}~\bibnamefont {Vanden-Eijnden}}\ and\ \bibinfo {author} {\bibfnamefont {F.~A.}\ \bibnamefont {Tal}},\ }\bibfield  {title} {\bibinfo {title} {Transition state theory: Variational formulation, dynamical corrections, and error estimates},\ }\href {https://doi.org/10.1063/1.2102898} {\bibfield  {journal} {\bibinfo  {journal} {J. Chem. Phys.}\ }\textbf {\bibinfo {volume} {123}},\ \bibinfo {pages} {184103} (\bibinfo {year} {2005})}\BibitemShut {NoStop}%
\bibitem [{\citenamefont {Krivov}(2018)}]{K18}%
  \BibitemOpen
  \bibfield  {author} {\bibinfo {author} {\bibfnamefont {S.~V.}\ \bibnamefont {Krivov}},\ }\bibfield  {title} {\bibinfo {title} {{Protein Folding Free Energy Landscape along the Committor - the Optimal Folding Coordinate}},\ }\href {https://doi.org/10.1021/acs.jctc.8b00101} {\bibfield  {journal} {\bibinfo  {journal} {J. Chem. Theory and Comput.}\ }\textbf {\bibinfo {volume} {14}},\ \bibinfo {pages} {3418} (\bibinfo {year} {2018})}\BibitemShut {NoStop}%
\bibitem [{\citenamefont {Roux}(2022)}]{roux2022transition}%
  \BibitemOpen
  \bibfield  {author} {\bibinfo {author} {\bibfnamefont {B.}~\bibnamefont {Roux}},\ }\bibfield  {title} {\bibinfo {title} {{Transition rate theory, spectral analysis, and reactive paths}},\ }\href {https://doi.org/10.1063/5.0084209} {\bibfield  {journal} {\bibinfo  {journal} {J. Chem. Phys}\ }\textbf {\bibinfo {volume} {156}},\ \bibinfo {pages} {134111} (\bibinfo {year} {2022})}\BibitemShut {NoStop}%
\bibitem [{\citenamefont {He}\ \emph {et~al.}(2022)\citenamefont {He}, \citenamefont {Chipot},\ and\ \citenamefont {Roux}}]{he_committor-consistent_2022}%
  \BibitemOpen
  \bibfield  {author} {\bibinfo {author} {\bibfnamefont {Z.}~\bibnamefont {He}}, \bibinfo {author} {\bibfnamefont {C.}~\bibnamefont {Chipot}},\ and\ \bibinfo {author} {\bibfnamefont {B.}~\bibnamefont {Roux}},\ }\bibfield  {title} {\bibinfo {title} {Committor-consistent variational string method},\ }\href {https://doi.org/10.1021/acs.jpclett.2c02529} {\bibfield  {journal} {\bibinfo  {journal} {J. Phys. Chem. Lett.}\ }\textbf {\bibinfo {volume} {13}},\ \bibinfo {pages} {9263} (\bibinfo {year} {2022})}\BibitemShut {NoStop}%
\bibitem [{\citenamefont {Chen}\ \emph {et~al.}(2023)\citenamefont {Chen}, \citenamefont {Roux},\ and\ \citenamefont {Chipot}}]{chen_discovering_2023}%
  \BibitemOpen
  \bibfield  {author} {\bibinfo {author} {\bibfnamefont {H.}~\bibnamefont {Chen}}, \bibinfo {author} {\bibfnamefont {B.}~\bibnamefont {Roux}},\ and\ \bibinfo {author} {\bibfnamefont {C.}~\bibnamefont {Chipot}},\ }\bibfield  {title} {\bibinfo {title} {Discovering reaction pathways, slow variables, and committor probabilities with machine learning},\ }\href {https://doi.org/10.1021/acs.jctc.3c00028} {\bibfield  {journal} {\bibinfo  {journal} {J. Chem. Theory Comput.}\ }\textbf {\bibinfo {volume} {19}},\ \bibinfo {pages} {4414} (\bibinfo {year} {2023})}\BibitemShut {NoStop}%
\bibitem [{\citenamefont {Berezhkovskii}\ and\ \citenamefont {Szabo}(2005)}]{berezhkovskii_one-dimensional_2005}%
  \BibitemOpen
  \bibfield  {author} {\bibinfo {author} {\bibfnamefont {A.}~\bibnamefont {Berezhkovskii}}\ and\ \bibinfo {author} {\bibfnamefont {A.}~\bibnamefont {Szabo}},\ }\bibfield  {title} {\bibinfo {title} {One-dimensional reaction coordinates for diffusive activated rate processes in many dimensions},\ }\href {https://doi.org/10.1063/1.1818091} {\bibfield  {journal} {\bibinfo  {journal} {J. Chem. Phys.}\ }\textbf {\bibinfo {volume} {122}},\ \bibinfo {pages} {014503} (\bibinfo {year} {2005})}\BibitemShut {NoStop}%
\bibitem [{\citenamefont {Louwerse}\ and\ \citenamefont {Sivak}(2022)}]{LS22}%
  \BibitemOpen
  \bibfield  {author} {\bibinfo {author} {\bibfnamefont {M.~D.}\ \bibnamefont {Louwerse}}\ and\ \bibinfo {author} {\bibfnamefont {D.~A.}\ \bibnamefont {Sivak}},\ }\bibfield  {title} {\bibinfo {title} {Information thermodynamics of the transition-path ensemble},\ }\href {https://doi.org/10.1103/PhysRevLett.128.170602} {\bibfield  {journal} {\bibinfo  {journal} {Phys. Rev. Lett.}\ }\textbf {\bibinfo {volume} {128}},\ \bibinfo {pages} {170602} (\bibinfo {year} {2022})}\BibitemShut {NoStop}%
\bibitem [{\citenamefont {Meg\'ias}\ \emph {et~al.}(2025)\citenamefont {Meg\'ias}, \citenamefont {{Contreras Arredondo}}, \citenamefont {Chen}, \citenamefont {Tang}, \citenamefont {Roux},\ and\ \citenamefont {Chipot}}]{MCCCRC25}%
  \BibitemOpen
  \bibfield  {author} {\bibinfo {author} {\bibfnamefont {A.}~\bibnamefont {Meg\'ias}}, \bibinfo {author} {\bibfnamefont {S.}~\bibnamefont {{Contreras Arredondo}}}, \bibinfo {author} {\bibfnamefont {C.~G.}\ \bibnamefont {Chen}}, \bibinfo {author} {\bibfnamefont {C.}~\bibnamefont {Tang}}, \bibinfo {author} {\bibfnamefont {B.}~\bibnamefont {Roux}},\ and\ \bibinfo {author} {\bibfnamefont {C.}~\bibnamefont {Chipot}},\ }\bibfield  {title} {\bibinfo {title} {Iterative variational learning of committor-consistent transition pathways using artificial neural networks},\ }\href {https://doi.org/10.1038/s43588-025-00828-3} {\bibfield  {journal} {\bibinfo  {journal} {Nat. Comput. Sci.}\ }\textbf {\bibinfo {volume} {5}},\ \bibinfo {pages} {592–602} (\bibinfo {year} {2025})}\BibitemShut {NoStop}%
\bibitem [{\citenamefont {Yan}\ and\ \citenamefont {Schlick}(2025)}]{YS25}%
  \BibitemOpen
  \bibfield  {author} {\bibinfo {author} {\bibfnamefont {S.}~\bibnamefont {Yan}}\ and\ \bibinfo {author} {\bibfnamefont {T.}~\bibnamefont {Schlick}},\ }\bibfield  {title} {\bibinfo {title} {{Heterogeneous and multiple conformational transition pathways between pseudoknots of the SARS-CoV-2 frameshift element}},\ }\href {https://doi.org/10.1073/pnas.2417479122} {\bibfield  {journal} {\bibinfo  {journal} {Proc. Natl. Acad. Sci. U.S.A.}\ }\textbf {\bibinfo {volume} {122}},\ \bibinfo {pages} {e2417479122} (\bibinfo {year} {2025})}\BibitemShut {NoStop}%
\bibitem [{\citenamefont {Meng}\ \emph {et~al.}(2016)\citenamefont {Meng}, \citenamefont {Shukla}, \citenamefont {Pande},\ and\ \citenamefont {Roux}}]{MSPR16}%
  \BibitemOpen
  \bibfield  {author} {\bibinfo {author} {\bibfnamefont {Y.}~\bibnamefont {Meng}}, \bibinfo {author} {\bibfnamefont {D.}~\bibnamefont {Shukla}}, \bibinfo {author} {\bibfnamefont {V.~S.}\ \bibnamefont {Pande}},\ and\ \bibinfo {author} {\bibfnamefont {B.}~\bibnamefont {Roux}},\ }\bibfield  {title} {\bibinfo {title} {Transition path theory analysis of c-src kinase activation},\ }\href {https://doi.org/10.1073/pnas.1602790113} {\bibfield  {journal} {\bibinfo  {journal} {Proc. Natl. Acad. Sci. U.S.A.}\ }\textbf {\bibinfo {volume} {113}},\ \bibinfo {pages} {9193} (\bibinfo {year} {2016})}\BibitemShut {NoStop}%
\bibitem [{\citenamefont {Barz}\ \emph {et~al.}(2019)\citenamefont {Barz}, \citenamefont {Rodner}, \citenamefont {Garcia},\ and\ \citenamefont {Denzler}}]{BRGD19}%
  \BibitemOpen
  \bibfield  {author} {\bibinfo {author} {\bibfnamefont {B.}~\bibnamefont {Barz}}, \bibinfo {author} {\bibfnamefont {E.}~\bibnamefont {Rodner}}, \bibinfo {author} {\bibfnamefont {Y.~G.}\ \bibnamefont {Garcia}},\ and\ \bibinfo {author} {\bibfnamefont {J.}~\bibnamefont {Denzler}},\ }\bibfield  {title} {\bibinfo {title} {Detecting regions of maximal divergence for spatio-temporal anomaly detection},\ }\href {https://doi.org/10.1109/TPAMI.2018.2823766} {\bibfield  {journal} {\bibinfo  {journal} {IEEE Trans. Pattern Anal. Mach. Intell.}\ }\textbf {\bibinfo {volume} {41}},\ \bibinfo {pages} {1088} (\bibinfo {year} {2019})}\BibitemShut {NoStop}%
\bibitem [{\citenamefont {Meg\'{\i}as}\ and\ \citenamefont {Santos}(2021)}]{MS21}%
  \BibitemOpen
  \bibfield  {author} {\bibinfo {author} {\bibfnamefont {A.}~\bibnamefont {Meg\'{\i}as}}\ and\ \bibinfo {author} {\bibfnamefont {A.}~\bibnamefont {Santos}},\ }\bibfield  {title} {\bibinfo {title} {{Hydrodynamics of granular gases of inelastic and rough hard disks or spheres. II. Stability analysis}},\ }\href {https://doi.org/10.1103/PhysRevE.104.034902} {\bibfield  {journal} {\bibinfo  {journal} {Phys. Rev. E}\ }\textbf {\bibinfo {volume} {104}},\ \bibinfo {pages} {034902} (\bibinfo {year} {2021})}\BibitemShut {NoStop}%
\bibitem [{\citenamefont {Harris}\ and\ \citenamefont {Roux}(2025)}]{harris2025membrane}%
  \BibitemOpen
  \bibfield  {author} {\bibinfo {author} {\bibfnamefont {J.}~\bibnamefont {Harris}}\ and\ \bibinfo {author} {\bibfnamefont {B.}~\bibnamefont {Roux}},\ }\bibfield  {title} {\bibinfo {title} {Membrane permeability of sucrose calculated from equilibrium time-correlation functions using molecular dynamics simulations with enhanced sampling},\ }\href {https://doi.org/10.1021/acs.jpcb.5c02905} {\bibfield  {journal} {\bibinfo  {journal} {J. Phys. Chem. B}\ }\textbf {\bibinfo {volume} {129}},\ \bibinfo {pages} {7172} (\bibinfo {year} {2025})}\BibitemShut {NoStop}%
\bibitem [{\citenamefont {Kumar}\ \emph {et~al.}(1992)\citenamefont {Kumar}, \citenamefont {Rosenberg}, \citenamefont {Bouzida}, \citenamefont {Swendsen},\ and\ \citenamefont {Kollman}}]{Kumar1992}%
  \BibitemOpen
  \bibfield  {author} {\bibinfo {author} {\bibfnamefont {S.}~\bibnamefont {Kumar}}, \bibinfo {author} {\bibfnamefont {J.~M.}\ \bibnamefont {Rosenberg}}, \bibinfo {author} {\bibfnamefont {D.}~\bibnamefont {Bouzida}}, \bibinfo {author} {\bibfnamefont {R.~H.}\ \bibnamefont {Swendsen}},\ and\ \bibinfo {author} {\bibfnamefont {P.~A.}\ \bibnamefont {Kollman}},\ }\bibfield  {title} {\bibinfo {title} {The weighted histogram analysis method for free-energy calculations on biomolecules. i. the method},\ }\href {https://doi.org/https://doi.org/10.1002/jcc.540130812} {\bibfield  {journal} {\bibinfo  {journal} {J. Comput. Chem.}\ }\textbf {\bibinfo {volume} {13}},\ \bibinfo {pages} {1011} (\bibinfo {year} {1992})}\BibitemShut {NoStop}%
\bibitem [{\citenamefont {Metzner}\ \emph {et~al.}(2006)\citenamefont {Metzner}, \citenamefont {Schütte},\ and\ \citenamefont {Vanden-Eijnden}}]{metzner_2006}%
  \BibitemOpen
  \bibfield  {author} {\bibinfo {author} {\bibfnamefont {P.}~\bibnamefont {Metzner}}, \bibinfo {author} {\bibfnamefont {C.}~\bibnamefont {Schütte}},\ and\ \bibinfo {author} {\bibfnamefont {E.}~\bibnamefont {Vanden-Eijnden}},\ }\bibfield  {title} {\bibinfo {title} {{Illustration of transition path theory on a collection of simple examples}},\ }\href {https://doi.org/10.1063/1.2335447} {\bibfield  {journal} {\bibinfo  {journal} {J. Chem. Phys.}\ }\textbf {\bibinfo {volume} {125}},\ \bibinfo {pages} {084110} (\bibinfo {year} {2006})}\BibitemShut {NoStop}%
\bibitem [{\citenamefont {Metzner}\ \emph {et~al.}(2009)\citenamefont {Metzner}, \citenamefont {Sch\"{u}tte},\ and\ \citenamefont {Vanden-Eijnden}}]{metzner_2008}%
  \BibitemOpen
  \bibfield  {author} {\bibinfo {author} {\bibfnamefont {P.}~\bibnamefont {Metzner}}, \bibinfo {author} {\bibfnamefont {C.}~\bibnamefont {Sch\"{u}tte}},\ and\ \bibinfo {author} {\bibfnamefont {E.}~\bibnamefont {Vanden-Eijnden}},\ }\bibfield  {title} {\bibinfo {title} {{Transition Path Theory for Markov Jump Processes}},\ }\href {https://doi.org/10.1137/070699500} {\bibfield  {journal} {\bibinfo  {journal} {Multiscale Model. Simul.}\ }\textbf {\bibinfo {volume} {7}},\ \bibinfo {pages} {1192} (\bibinfo {year} {2009})}\BibitemShut {NoStop}%
\bibitem [{\citenamefont {Rosso}\ \emph {et~al.}(2005)\citenamefont {Rosso}, \citenamefont {Abrams},\ and\ \citenamefont {Tuckerman}}]{RAT05}%
  \BibitemOpen
  \bibfield  {author} {\bibinfo {author} {\bibfnamefont {L.}~\bibnamefont {Rosso}}, \bibinfo {author} {\bibfnamefont {J.~B.}\ \bibnamefont {Abrams}},\ and\ \bibinfo {author} {\bibfnamefont {M.~E.}\ \bibnamefont {Tuckerman}},\ }\bibfield  {title} {\bibinfo {title} {Mapping the backbone dihedral free-energy surfaces in small peptides in solution using adiabatic free-energy dynamics},\ }\href {https://doi.org/10.1021/jp045399i} {\bibfield  {journal} {\bibinfo  {journal} {J. Phys. Chem. B}\ }\textbf {\bibinfo {volume} {109}},\ \bibinfo {pages} {4162} (\bibinfo {year} {2005})}\BibitemShut {NoStop}%
\bibitem [{\citenamefont {H{\'e}nin}\ \emph {et~al.}(2010)\citenamefont {H{\'e}nin}, \citenamefont {Fiorin}, \citenamefont {Chipot},\ and\ \citenamefont {Klein}}]{HFCK10}%
  \BibitemOpen
  \bibfield  {author} {\bibinfo {author} {\bibfnamefont {J.}~\bibnamefont {H{\'e}nin}}, \bibinfo {author} {\bibfnamefont {G.}~\bibnamefont {Fiorin}}, \bibinfo {author} {\bibfnamefont {C.}~\bibnamefont {Chipot}},\ and\ \bibinfo {author} {\bibfnamefont {M.~L.}\ \bibnamefont {Klein}},\ }\bibfield  {title} {\bibinfo {title} {Exploring multidimensional free energy landscapes using time-dependent biases on collective variables},\ }\href {https://doi.org/10.1021/ct9004432} {\bibfield  {journal} {\bibinfo  {journal} {J. Chem. Theory Comput.}\ }\textbf {\bibinfo {volume} {6}},\ \bibinfo {pages} {35} (\bibinfo {year} {2010})}\BibitemShut {NoStop}%
\bibitem [{\citenamefont {Bolhuis}\ \emph {et~al.}(2000)\citenamefont {Bolhuis}, \citenamefont {Dellago},\ and\ \citenamefont {Chandler}}]{bolhuis_reaction_2000}%
  \BibitemOpen
  \bibfield  {author} {\bibinfo {author} {\bibfnamefont {P.~G.}\ \bibnamefont {Bolhuis}}, \bibinfo {author} {\bibfnamefont {C.}~\bibnamefont {Dellago}},\ and\ \bibinfo {author} {\bibfnamefont {D.}~\bibnamefont {Chandler}},\ }\bibfield  {title} {\bibinfo {title} {Reaction coordinates of biomolecular isomerization},\ }\href {https://doi.org/10.1073/pnas.100127697} {\bibfield  {journal} {\bibinfo  {journal} {Proc. Natl. Acad. Sci. U.S.A.}\ }\textbf {\bibinfo {volume} {97}},\ \bibinfo {pages} {5877} (\bibinfo {year} {2000})}\BibitemShut {NoStop}%
\bibitem [{\citenamefont {Kang}\ \emph {et~al.}(2024)\citenamefont {Kang}, \citenamefont {Trizio},\ and\ \citenamefont {Parrinello}}]{kang2024Parrinello}%
  \BibitemOpen
  \bibfield  {author} {\bibinfo {author} {\bibfnamefont {P.}~\bibnamefont {Kang}}, \bibinfo {author} {\bibfnamefont {E.}~\bibnamefont {Trizio}},\ and\ \bibinfo {author} {\bibfnamefont {M.}~\bibnamefont {Parrinello}},\ }\bibfield  {title} {\bibinfo {title} {Computing the committor with the committor to study the transition state ensemble},\ }\href {https://doi.org/10.1038/s43588-024-00645-0} {\bibfield  {journal} {\bibinfo  {journal} {Nat. Comput. Sci.}\ }\textbf {\bibinfo {volume} {4}},\ \bibinfo {pages} {451–460} (\bibinfo {year} {2024})}\BibitemShut {NoStop}%
\bibitem [{\citenamefont {Trizio}\ \emph {et~al.}(2025)\citenamefont {Trizio}, \citenamefont {Kang},\ and\ \citenamefont {Parrinello}}]{Trizio2025}%
  \BibitemOpen
  \bibfield  {author} {\bibinfo {author} {\bibfnamefont {E.}~\bibnamefont {Trizio}}, \bibinfo {author} {\bibfnamefont {P.}~\bibnamefont {Kang}},\ and\ \bibinfo {author} {\bibfnamefont {M.}~\bibnamefont {Parrinello}},\ }\bibfield  {title} {\bibinfo {title} {Everything everywhere all at once: a probability-based enhanced sampling approach to rare events},\ }\href {https://doi.org/10.1038/s43588-025-00799-5} {\bibfield  {journal} {\bibinfo  {journal} {Nat. Comput. Sci.}\ }\textbf {\bibinfo {volume} {5}},\ \bibinfo {pages} {582–591} (\bibinfo {year} {2025})}\BibitemShut {NoStop}%
\bibitem [{\citenamefont {Lee}\ \emph {et~al.}(2017)\citenamefont {Lee}, \citenamefont {Lee}, \citenamefont {Joung}, \citenamefont {Lee},\ and\ \citenamefont {Brooks}}]{Lee2017}%
  \BibitemOpen
  \bibfield  {author} {\bibinfo {author} {\bibfnamefont {J.}~\bibnamefont {Lee}}, \bibinfo {author} {\bibfnamefont {I.-H.}\ \bibnamefont {Lee}}, \bibinfo {author} {\bibfnamefont {I.}~\bibnamefont {Joung}}, \bibinfo {author} {\bibfnamefont {J.}~\bibnamefont {Lee}},\ and\ \bibinfo {author} {\bibfnamefont {B.~R.}\ \bibnamefont {Brooks}},\ }\bibfield  {title} {\bibinfo {title} {Finding multiple reaction pathways via global optimization of action},\ }\href {https://doi.org/10.1038/ncomms15443} {\bibfield  {journal} {\bibinfo  {journal} {Nat. Commun.}\ }\textbf {\bibinfo {volume} {8}},\ \bibinfo {pages} {15443} (\bibinfo {year} {2017})}\BibitemShut {NoStop}%
\bibitem [{\citenamefont {Tiwary}\ and\ \citenamefont {Berne}(2017)}]{Tiwary2017Predicting}%
  \BibitemOpen
  \bibfield  {author} {\bibinfo {author} {\bibfnamefont {P.}~\bibnamefont {Tiwary}}\ and\ \bibinfo {author} {\bibfnamefont {B.~J.}\ \bibnamefont {Berne}},\ }\bibfield  {title} {\bibinfo {title} {{Predicting reaction coordinates in energy landscapes with diffusion anisotropy}},\ }\href {https://doi.org/10.1063/1.4983727} {\bibfield  {journal} {\bibinfo  {journal} {J. Chem. Phys.}\ }\textbf {\bibinfo {volume} {147}},\ \bibinfo {pages} {152701} (\bibinfo {year} {2017})}\BibitemShut {NoStop}%
\bibitem [{\citenamefont {Chen}\ \emph {et~al.}(2022)\citenamefont {Chen}, \citenamefont {Ogden}, \citenamefont {Pant}, \citenamefont {Cai}, \citenamefont {Tajkhorshid}, \citenamefont {Moradi}, \citenamefont {Roux},\ and\ \citenamefont {Chipot}}]{chen_companion_2022}%
  \BibitemOpen
  \bibfield  {author} {\bibinfo {author} {\bibfnamefont {H.}~\bibnamefont {Chen}}, \bibinfo {author} {\bibfnamefont {D.}~\bibnamefont {Ogden}}, \bibinfo {author} {\bibfnamefont {S.}~\bibnamefont {Pant}}, \bibinfo {author} {\bibfnamefont {W.}~\bibnamefont {Cai}}, \bibinfo {author} {\bibfnamefont {E.}~\bibnamefont {Tajkhorshid}}, \bibinfo {author} {\bibfnamefont {M.}~\bibnamefont {Moradi}}, \bibinfo {author} {\bibfnamefont {B.}~\bibnamefont {Roux}},\ and\ \bibinfo {author} {\bibfnamefont {C.}~\bibnamefont {Chipot}},\ }\bibfield  {title} {\bibinfo {title} {A companion guide to the string method with swarms of trajectories: Characterization, performance, and pitfalls},\ }\href {https://doi.org/10.1021/acs.jctc.1c01049} {\bibfield  {journal} {\bibinfo  {journal} {J. Chem. Theory Comput.}\ }\textbf {\bibinfo {volume} {18}},\ \bibinfo {pages} {1406} (\bibinfo {year} {2022})}\BibitemShut {NoStop}%
\bibitem [{\citenamefont {Fu}\ \emph {et~al.}(2019)\citenamefont {Fu}, \citenamefont {Shao}, \citenamefont {Cai},\ and\ \citenamefont {Chipot}}]{fu_taming_2019}%
  \BibitemOpen
  \bibfield  {author} {\bibinfo {author} {\bibfnamefont {H.}~\bibnamefont {Fu}}, \bibinfo {author} {\bibfnamefont {X.}~\bibnamefont {Shao}}, \bibinfo {author} {\bibfnamefont {W.}~\bibnamefont {Cai}},\ and\ \bibinfo {author} {\bibfnamefont {C.}~\bibnamefont {Chipot}},\ }\bibfield  {title} {\bibinfo {title} {Taming rugged free energy landscapes using an average force},\ }\href {https://doi.org/10.1021/acs.accounts.9b00473} {\bibfield  {journal} {\bibinfo  {journal} {Acc. Chem. Res.}\ }\textbf {\bibinfo {volume} {52}},\ \bibinfo {pages} {3254} (\bibinfo {year} {2019})}\BibitemShut {NoStop}%
\end{thebibliography}
\end{document}